\documentclass[
aps
,groupedaddress
,nofootinbib
,showkeys
,showpacs
,amsfonts
,eqsecnum
,superscriptaddress
]{revtex4}

\usepackage{bm}
\usepackage{amsmath}
\newcommand{\tr}{{\rm tr \,}}
\newcommand{\Tr}{{\rm Tr}}

\newcommand{\eq}[1]{eq.~(\ref{eq:#1})}
\newcommand{\eqs}[1]{eqs.~(\ref{eq:#1})}
\newcommand{\Eq}[1]{Eq.~(\ref{eq:#1})}

\newcommand{\ignore}[1]{}

\newcommand{\D}{D}
\newcommand{\R}{\mathbb{R}}
\newcommand{\Z}{\mathbb{Z}}

\newcommand{\va}{{\bm a}}
\newcommand{\vx}{{\bm x}}
\newcommand{\vp}{{\bm p}}
\newcommand{\vk}{{\bm k}}
\newcommand{\vD}{{\bm D}}

\newcommand{\dimst}{d}
\newcommand{\dims}{d-1}

\begin{document} 

\title{Derivative expansion of the heat kernel at finite temperature}

\author{F. J. Moral-G{\'a}mez}
\affiliation{Departamento de F{\'\i}sica At{\'o}mica, Molecular y Nuclear,
  Universidad de Granada, E-18071 Granada, Spain}

\author{L. L. Salcedo}
\email{salcedo@ugr.es}
\affiliation{Departamento de F{\'\i}sica At{\'o}mica, Molecular y Nuclear,
  Universidad de Granada, E-18071 Granada, Spain}
\affiliation{Instituto Carlos I de F{\'\i}sica Te\'orica y Computacional,
  Universidad de Granada, E-18071 Granada, Spain}

\date{\today}

\begin{abstract}
The method of covariant symbols of Pletnev and Banin is extended to
space-times with topology $\R^n\times S^1\times \cdots \times S^1$.  By means
of this tool, we obtain explicit formulas for the diagonal matrix elements and
the trace of the heat kernel at finite temperature to fourth order in a strict
covariant derivative expansion. The role of the Polyakov loop is
emphasized. Chan's formula for the effective action to one loop is similarly
extended. The expressions obtained apply formally to a larger class of spaces,
$h$-spaces, with an arbitrary weight function $h(p)$ in the integration over
the momentum of the loop.
\end{abstract}



\maketitle
\tableofcontents

\section{Introduction}
\label{sec:1}

Among other uses, the heat kernel \cite{Vassilevich:2003xt} is a tool to deal
with one-loop effective actions in quantum field theory.  The effective
action, the trace of the logarithm of the fluctuation operator
\cite{Ramond:1989yd}, suffers from ultraviolet divergences, as well as
many-valuation and anomalies. As noted in \cite{Schwinger:1951nm} the heat
kernel has the virtue of being one-valued, free from ultraviolet divergences
and gauge invariant.

The heat kernel finds a number of applications: study of spectral densities of
Klein-Gordon operators, proof of index theorems
\cite{Gilkey:1975iq,Atiyah:1973ad}, to compute the $\zeta$-function
\cite{Hawking:1977ja} and the anomalies of Dirac operators
\cite{Fujikawa:1980eg}, to deal with chiral gauge theories \cite{Ball:1989xg}
and models of QCD \cite{Bijnens:1996ww}, to the Casimir effect
\cite{Bordag:2001qi}, to compute black hole entropies \cite{Callan:1994py},
etc.

Except in very particular manifolds, the heat kernel is expressed by means of
asymptotic expansions. The 
Seeley-DeWitt expansion \cite{Dewitt:1975ys,Seeley:1967ea},
is in powers of the proper time, and is available
to rather high orders in
several setups, including curved spaces with and without boundary, and in
presence of non Abelian gauge fields and non Abelian scalar fields, using
different methods
\cite{Ball:1989xg,Bel'kov:1996tn,vandeVen:1998pf,Moss:1999wq,%
  Fliegner:1998rk,Avramidi:1991je,Gusynin:1989ky,Elizalde:1994bk,%
  Vassilevich:2003xt}.

To study quantum field theory at finite temperature one can use the imaginary
time formalism, with compactified Euclidean time
\cite{Matsubara:1955ws,Landsman:1986uw}. This introduces a modification in the
heat kernel coefficients. Early attempts to compute those coefficients were
made in \cite{Boschi-Filho:1992ah,Xu:1993qi}. However, ad hoc assumptions made
in those calculations (essentially what we call the quenched approximation
below) lead to expressions in conflict with explicit results derived for
particular settings \cite{Actor:2000hc,Gies:1998vt}. The first systematic, and
fully gauge covariant, calculation of the heat kernel at finite temperature
was presented in \cite{Megias:2002vr,Megias:2003ui}. There it was found that,
besides the usual covariant derivatives, the Polyakov loop, $\Omega(x)$, was
also present in the expressions (consistently with
\cite{Actor:2000hc,Gies:1998vt}). This is to be expected, since the Polyakov
loop is the other natural gauge covariant construction allowed at finite
temperature. This is not just a technical nicety, in fact, nowadays the
gluonic Polyakov loop in QCD at finite temperature plays a prominent role as a
relevant order parameter of confinement in the very successful
Polyakov--Nambu--Jona-Lasinio models
\cite{Fukushima:2003fw,Megias:2004hj,Ratti:2005jh}. The Polyakov loop appears
automatically in any gauge covariant computation at finite temperature, and
solves long standing paradoxes related to gauge invariance due to naive
perturbative expansions
\cite{Dunne:1996yb,GarciaRecio:2000gt,Salcedo:2002pr}. Moreover it is the
only way a chemical potential could appear in the effective action.  Indeed,
the chemical potential is obtained by the shift $A_0(x) \to A_0(x)-\mu$, where
$\mu$ is a constant real c-number. This has no effect in $[D_0,~]$, but it
shows through the Polyakov loop dependence due to $\Omega(x)\to e^{\beta
  \mu}\Omega(x)$ \cite{GarciaRecio:2000gt,Gies:1998vt}.

The results of \cite{Megias:2002vr,Megias:2003ui} refer to the usual heat
kernel expansion. That is, the coefficients are classified according to the
dimension of the operators they carry (this classification holds at zero or
finite temperature, and at zero temperature is equivalent to an expansion in
the powers of the proper time). In \cite{Salcedo:2004yh} an expansion of the
(zero temperature) heat kernel based on the number of covariant derivatives
was carried out. This is a resummation of the usual expansion in which each
coefficient has a fixed number of covariant derivatives but any number of
scalar fields. The extension to curved space-time was made in
\cite{Salcedo:2007bt}. In the present work we compute, for the first time, the
heat kernel at finite temperature within the covariant derivative expansion.

The results for the heat kernel at finite temperature of
\cite{Megias:2002vr,Megias:2003ui} were obtained using a rather cumbersome
method. Essentially it was a mixture of (already known) zero temperature
coefficients for the spatial covariant derivatives plus the method of symbols
\cite{Nepomechie:1984wt,Salcedo:1996qy} for the covariant time derivative. In
this approach some work is required to bring the expression to a manifestly
gauge covariant form, involving the Polyakov loop. This is largely improved in
the present paper. The new idea presented here is based on extending the
method of covariant symbols, introduced by Pletnev and Banin
\cite{Pletnev:1998yu}, to the finite temperature case. The original method was
devised for zero temperature, and so it assumed a continuous frequency
variable. We adapt here the method so that it applies also for the discrete
Matsubara frequencies. The Polyakov loop is accommodated in a natural way in
the new approach. By means of this new technique, the calculation of the heat
kernel at finite temperature, or other quantities like the effective action,
can be done with manifest gauge covariance at each step. The method applies to
general pseudo-differential operators.

In loop momentum integrals, the spatial components are continuous, but the
frequency becomes discrete as a consequence of periodicity. This is equivalent
to introducing a weight function in momentum space which consists of a family
of Dirac deltas with support at the Matsubara frequencies. Here we find the
remarkable result that much of the formalism goes through also for completely
general weight functions $h(p)$ in momentum space. This allows to obtain
expressions which look Lorentz covariant (prior to momentum integration).  The
finite temperature case can be obtained from the generic one by replacing
$h(p)$ by its Matsubara version. As a third contribution of this work, we
adapt Chan's formula for the effective action \cite{Chan:1986jq} to such
$h$-spaces, and so in particular to finite temperature. (This automatically
implies the corresponding result for the heat kernel.) The existence of Chan's
form in such general setting is far from obvious a priori since the original
construction by Chan relied heavily on integration by parts and averages in
momentum space. These tools are not available in the presence of a generic
weight $h(p)$.

The paper is organized as follows. In Section \ref{sec:2} we make a summary of
previous results and techniques and develop the new method of covariant
symbols valid at finite temperature, \eq{2.36}. In Section \ref{sec:3} we
present explicit results for the strict covariant derivative expansion of the
heat kernel at finite temperature, to third order for the diagonal matrix
elements, \eq{3.26}, and to fourth order for the trace, eqs. (\ref{eq:3.31})
and (\ref{eq:3.33}). Non stationary and non Abelian configurations are
assumed throughout. In Section \ref{sec:4} we extend the gauge covariant
technique to $h$-spaces and use it to obtain the very compact Chan's form of
the effective action, eqs. (\ref{eq:4.22}) and (\ref{eq:4.23}). In Section
\ref{sec:5} we summarize our conclusions.  Some auxiliary material and results
are given in the appendices.

\section{Method of symbols}
\label{sec:2}

\subsection{General considerations}
\label{sec:2A}

Let us consider a theory of scalar fields in $\dimst$-dimensional Euclidean
flat space-time, coupled to external fields, including gauge fields. Typically
\begin{equation}
{\mathcal L}(x) = -\phi(x)^\dagger K  \phi(x)
,
\qquad
K = D^2+X(x)
,
\qquad
D_\mu = \partial_\mu + A_\mu(x)
.
\label{eq:2.1}
\end{equation}
The external fields $X(x)$ and $A_\mu(x)$ are matrices in internal space in
general. For concreteness we assume that $\phi(x)$ transforms in the
fundamental representation of the gauge group, $\phi(x)\to U^{-1}(x)\phi(x)$.

The corresponding partition function and effective action
are
\begin{equation}
Z= \int {\cal D}\phi^\dagger{\cal D}\phi\,e^{-\int d^{\dimst}x\,{\mathcal L}(x)}
=
e^{-{\Gamma}}
,
\qquad
\Gamma = \Tr\log K
.
\label{eq:2.3}
\end{equation}
The effective action $\Gamma$ is a functional of the external fields and
diagrammatically $\Tr\log$ corresponds to adding one-loop graphs with the
field $\phi$ running in the loop and any number of external legs attached to
it.

The operation $\Tr$ can be expressed as a trace on a single-particle
Hilbert space where $K$ acts. This Hilbert space includes
space-time and also internal degrees of freedom:
\begin{equation}
\Gamma = \int d^{\dimst}x\,\tr\langle x|\log K|x\rangle
,
\end{equation}
$|x\rangle$ is a basis of the space-time sector, 
\begin{equation}
\langle x|x^\prime\rangle
=\delta(x-x^\prime)
,
\qquad
\hat{x}_\mu|x\rangle =  x_\mu|x\rangle
,
\end{equation}
and $\tr$ refers to the internal degrees of freedom. Likewise, under a
variation of the gauge fields and the scalar field, one obtains the current
and density,
\begin{eqnarray}
\delta\Gamma &=& \int d^{\dimst}x\,\tr(\mathcal{J}_\mu(x)\,\delta A_\mu(x)
+
\mathcal{D}(x)\,\delta X(x))
,
\nonumber \\
\mathcal{J}_\mu(x) &=& \langle x|\{K^{-1},D_\mu\}|x\rangle
,
\qquad
\mathcal{D}(x) = \langle x|K^{-1}|x\rangle
.
\end{eqnarray}

These examples, as well as the heat kernel, $\exp(\tau K)$, to be
considered later, illustrate the need for computing diagonal matrix elements
of pseudo-differential operators. Taking coincident points amounts to
integrate over the momentum of the loop.

In view of the above, we consider a generic pseudo-differential operator 
\begin{equation}
\hat{f}=f(D,X) ,
\end{equation}
constructed with the covariant derivative $D_\mu$ and other fields,
$X(x)$. These external field are bosonic. The quantum field running in the
loop may be bosonic or fermionic.  Under a gauge transformation, $D_\mu \to
U^{-1} D_\mu U$, $X \to U^{-1} X U$, the diagonal matrix
elements transform covariantly, $\langle x| f(D,X)|x\rangle \to
U^{-1}(x)\langle x| f(D,X)|x\rangle U(x)$.

Our goal is to address the computation of  the diagonal matrix elements of the
pseudo-differential operator, $\langle x| f(D,X)|x\rangle$, and its trace, in
a gauge covariant setting valid at zero or finite temperature.

\subsubsection{Covariant expansions at zero temperature}

In general the diagonal matrix element cannot be expressed in closed form.  At
zero temperature, a typical expansion to be applied is one based in powers of
$D_\mu$ and of $X(x)$. This produces an expansion in terms of local gauge
covariant operators
\begin{equation}
\langle x| f(D,X)|x\rangle
=
\sum_\lambda g_\lambda \mathcal{O}_\lambda(x)
\label{eq:2.7a}
.
\end{equation}
Here $\mathcal{O}_\lambda(x)$ includes all possible local gauge covariant
operators constructed with $D_\mu$ and $X$. That is, with $X$, with the field
strength tensor,
\begin{equation}
F_{\mu\nu}=[D_\mu,D_\nu]
,
\end{equation}
and with their covariant derivatives.  The coupling constants $g_\lambda$
depend on the concrete operator $\hat{f}$. Often the terms are organized by
dimensional counting in subsets of operators with a common dimension.
An example is the standard heat kernel
expansion
\begin{equation}
\langle x| e^{\tau K} |x\rangle
=
\frac{1}{(4\pi\tau)^{\dimst/2}}
\left(
1+\tau X + 
\tau^2 \left(
\frac{1}{2}X^2+ \frac{1}{6}X_{\mu\mu} 
+ \frac{1}{12}F_{\mu\nu}^2
\right)
+\cdots
\right)
.
\label{eq:2.9}
\end{equation}
We indicate covariant derivatives using the convention\footnote{Here and
  elsewhere $Y$ denotes a generic operator.}
\begin{equation}
Y_{\mu_1\mu_2\ldots\mu_n} = [D_{\mu_1},Y_{\mu_2\ldots\mu_n }]
,
\end{equation}
for any operator $Y_I$ with a (possibly empty) ordered set of Lorentz indices
$I$. So, for instance, $F_{\alpha\mu\nu}= [D_\alpha,F_{\mu\nu}]$ and
$X_{\alpha\beta}= [D_\alpha,[D_\beta,X]]$.

Another expansion, which is the subject of this work, is the {\em covariant
derivative expansion}, which is a resummation of the previous one: at a given
order the number of $D_\mu$ is fixed while there can be any number of $X$. For
Abelian $X$ this is just of the form
\begin{equation}
\langle x| f(D,X) |x\rangle
=
\sum_\lambda f_\lambda(X(x)) \, \mathcal{O}_\lambda(x)
,
\end{equation}
where now $\mathcal{O}_\lambda(x)$ contains only $X$ with derivatives,
and $f_\lambda(X(x))$ is a generic function of $X$. In the more general case
of non Abelian fields, one can still express the expansion by means of labeled
operators \cite{Salcedo:2004yh,Feynman:1951gn}:
\begin{equation}
\langle x| f(D,X) |x\rangle = \sum_\lambda f_\lambda(X_1(x),\ldots,X_n(x))
\,\mathcal{O}_\lambda(x)
.
\label{eq:2.10}
\end{equation}
The idea is that $\mathcal{O}_\lambda(x)$ is the product of $n-1$ local
covariant blocks and the $i$-th copy of $X$, $X_i$, is meant to act between
the $(i-1)$-th and the $i$-th block. For instance
\begin{equation}
\int_0^s e^{t X} F^2_{\mu\nu} e^{(s-t) X} \,dt
=
\int_0^s e^{t X_1}  e^{(s-t) X_2} \,dt F^2_{\mu\nu}
=
\frac{e^{s X_1}- e^{s X_2}}{X_1-X_2} \, F^2_{\mu\nu}
.
\end{equation}
Here $X_1$ is $X$ acting at the left of $F^2_{\mu\nu}$ and $X_2$ is $X$ acting
at the right. Note that the labeled operators $X_i$ can be treated as
c-numbers since $X_1X_2=X_2X_1$.

As an example, all the terms of the heat kernel with precisely one
$X_{\mu\mu}$ can be collected in the form \cite{Salcedo:2004yh}
\begin{equation}
\langle x| e^{\tau K} |x\rangle
=
\frac{1}{(4\pi\tau)^{\dimst/2}}
\left(
\cdots + 
 \left(
\frac{e^{\tau X_1}+e^{\tau X_2}}{(X_1-X_2)^2}
-\frac{2}{\tau}\frac{e^{\tau X_1}-e^{\tau X_2}}{(X_1-X_2)^3}
\right)
X_{\mu\mu} 
+\cdots
\right)
.
\end{equation}
Expanding in powers of $X_i$ gives back the standard heat-kernel
expansion:
\begin{eqnarray}
&&
\left(
\frac{e^{\tau X_1}+e^{\tau X_2}}{(X_1-X_2)^2}
-\frac{2}{\tau}\frac{e^{\tau X_1}-e^{\tau X_2}}{(X_1-X_2)^3}
\right)
X_{\mu\mu}
=
\left(\frac{\tau^2}{6}
+\frac{\tau^3}{12} (X_1+X_2)
+ \cdots
\right)X_{\mu\mu}
\nonumber\\
&&
=
\frac{\tau^2}{6}X_{\mu\mu}
+\frac{\tau^3}{12} 
\{ X, X_{\mu\mu} \}
 + \cdots
.
\end{eqnarray}

\subsubsection{Covariant expansions at finite temperature}

At finite temperature the space-time is $\R^{\dims}\times S^1$, within the
imaginary time formalism \cite{Matsubara:1955ws,Landsman:1986uw}. The quantum
field may be bosonic or fermionic, being respectively periodic or antiperiodic
in time with period $\beta=1/T$, where $T$ is the temperature.  The external
fields $A_\mu(x)$ and $X(x)$ are bosonic and hence periodic. The gauge
transformations, $U(x)$, are also periodic.

The expansions in \eq{2.7a} and (\ref{eq:2.10}) refer to zero temperature and
they have to be modified at finite temperature. In fact, at finite temperature
there are two gauge covariant constructions with the operator $D_0$, namely,
the covariant derivative $[D_0,~]$ and the Polyakov loop,
\begin{equation}
\Omega(x) =
 Pe^{-\int_{x_0}^{x_0+\beta} A_0(\vx,t)\,dt}
.
\end{equation}
The Polyakov loop here is not traced, it is a matrix in internal space and $P$
refers to path ordered product. Also, the integral starts at $x_0$ rather than
zero.  The Polyakov loop so defined is gauge covariant at $x$:
\begin{equation}
\Omega(x) \to U^{-1}(x) \Omega(x) U(x)
.
\end{equation}
$\Omega(x)$ is also periodic in $x_0$. In practical terms, $\Omega(x)$ behaves
as a local field. This operator appears through $D_0$ due to the relation
\cite{Megias:2002vr}
\begin{equation}
e^{-\beta D_0} = \Omega(x)
.
\end{equation}
The easiest way to show this is by going to a gauge where $A_0(x)$ is time
independent. In such a gauge $\Omega(x)= e^{-\beta A_0(\vx)}$, while
$e^{-\beta D_0}=e^{-\beta \partial_0}e^{-\beta A_0}$. But $e^{-\beta
  \partial_0}=1$ due to periodicity. The equality holds in any gauge since the
two operators $e^{-\beta D_0}$ and $\Omega(x)$ transform in the same way under
gauge transformations. Hence, although formally $\exp(-\beta D_0)$ would be a
pseudo-differential operator ($D_0$ being a differential operator), actually
it is just a multiplicative operator.\footnote{Multiplicative operators will
  be important in what follows. By multiplicative operators we mean zeroth
  order differential operators with respect to $x$. That is, operators which
  may contain $\hat{x}_\mu$ but not $\partial_\mu$. They can be matrices in
  internal space. Thus they are in one to one correspondence with ordinary
  matrix-valued functions of $x$.}

The two gauge covariant constructions, $[D_\mu,~]$ and $\Omega(x)$, appear at
finite temperature.  The heat kernel-like expansion (expansion in powers of
$D_\mu$ and $X$) in \eq{2.7a} is modified at finite temperature to

\begin{equation}
\langle x| f(D,X)|x\rangle
=
\sum_\lambda g_\lambda(\pm\Omega(x);T)\, \mathcal{O}_\lambda(x)
\label{eq:2.7b}
.
\end{equation}
Here $\mathcal{O}_\lambda(x)$ are still arbitrary local gauge covariant
operators constructed with $X$ and $[D_\mu,~]$. On the other hand
$g_\lambda(\pm\Omega(x);T)$ are functions of the Polyakov loop and the
temperature, determined by the pseudo-differential operator $\hat{f}$.  The
$\pm$ refers to the two cases of bosonic or fermionic quantum field,
respectively.

Note that, in general, $\Omega(x)$ does not commute with the local
operators. We have chosen to put all the dependence on the Polyakov loop at
the left. This can be done due to the identity \cite{Megias:2003ui}
\begin{equation}
[\mathcal{O},g(\Omega)] 
= \sum_{n=1}^\infty
\frac{i^n}{n!} g_n(\Omega)\hat{D}_0^n\mathcal{O}
\label{eq:2.18}
,
\end{equation}
where $g_n(\Omega)$ is just the $n$-th derivative of $g(\Omega)$ as a function
of the variable $iT\log(\Omega)$ and $\hat{D}_0=[D_0,~]$.

For instance, the expansion in \eq{2.7b} has been computed for the heat kernel
through operators of dimension 6 in \cite{Megias:2002vr,Megias:2003ui}:
\begin{equation}
\langle x| e^{\tau K} |x\rangle
=
\frac{1}{(4\pi\tau)^{\dimst/2}}
\left(
\xi_0 +  \tau \xi_0 X +
\cdots
\right)
,
\qquad
\xi_0 = \sum_{k\in\Z} (\pm\Omega)^k e^{-k^2\beta^2/4\tau}
.
\end{equation}
At zero temperature $\xi_0=1$ and this expression reduces to that in \eq{2.9}.

For the derivative expansion at finite temperature, one can write
\begin{equation}
\langle x| f(D,X) |x\rangle
=
\sum_\lambda f_\lambda(\pm\Omega,X_1,\ldots,X_n;T)\, \mathcal{O}_\lambda
,
\label{eq:2.7c}
\end{equation}
with the $\Omega$'s at the left of all $X$'s and $\mathcal{O}_\lambda$, and
$X_i$ is inserted between the $(i-1)$-th and the $i$-th blocks of
$\mathcal{O}_\lambda$, as before. (Recall that $\mathcal{O}_\lambda$ contains
only operators $X$ with derivatives. $X$ without derivatives go into
$f_\lambda$.)

The functions $f_\lambda$ in \eq{2.7c} are well defined but have not been
computed yet even for the heat kernel. This is a goal of this work.

\subsubsection{Countings at zero and finite temperature}

Before closing this section it is important to note that the counting of a term
either by its dimension, or by its number of derivatives, is not as clean at
finite temperature as it was at zero temperature. Indeed, to unambiguously
classify a term by its (scale) dimension at zero temperature one can
introduce a bookkeeping parameter $\lambda$ in the external fields, as
$\lambda A_\mu(\lambda x)$ and $\lambda^\alpha X(\lambda x)$ (being $\alpha$
the dimension of $X$, $\alpha=2$ in the example of \eq{2.1}).  In this way an
operator $\mathcal{O}_n$ of dimension $\gamma$ will be tagged by a factor
$\lambda^\gamma$. At finite temperature the number of $X$'s and $[\vD,~]$ can
still be counted by a bookkeeping parameter, but the method fails for $D_0$
because a dilation in the time direction is not consistent with periodicity of
the external fields.  Of course, this is related to the presence of discrete
values for $p_0$ and to the presence of $\Omega(x)$ in addition to $[D_0,~]$.

At finite temperature there is no bookkeeping parameter to fix the order of a
term in the dimensional expansion, and so the order is undefined or it may
look different depending on how the term is written. To sort out this problem,
we take the prescription of defining the counting {\em after the term has been
  written with all $\Omega(x)$ at the left}.  With this prescription the order
can be defined without ambiguity (see appendix \ref{app:A}). We take
$\Omega(x)$ to be of dimension zero. As before $X(x)$ has dimension $\alpha$,
$[D_\mu,~]$ has dimension one and $F_{\mu\nu}$ has dimension two. So for
instance, the operator $\Omega(x) X(x)$ carries dimension $\alpha$, whereas
(using \eq{2.18})
\begin{equation}
[X(x),\Omega(x)] =  \beta\Omega(x) [D_0,X(x)] + \cdots
\end{equation}
carries {\em leading} dimension $\alpha+1$ but is not homogeneous in this
counting. As usual, we will consider the leading order as the order of a non
homogeneous term.

Everything is similar for the derivative expansion. In this case the zero
temperature counting comes from $\lambda A_\mu(\lambda x)$ and $X(\lambda x)$.
At finite temperature, the term is written with $\Omega(x)$ at the left and
then $\Omega$ and $X$ count as order zero, $[D_\mu,~]$ as order one and
$F_{\mu\nu}(x)$ as order two. For instance, the operator $\Omega(x) X(x)$ is of
order zero whereas $[X(x),\Omega(x)]$ is of order one.

The situation for traced terms at finite temperature is more involved due to
the trace cyclic property. To define the order of an expression in this case,
the natural prescription is to consider all possible ways to write it and
select the one with highest leading order as the true order of the expression.
For instance, using the property $[D_0,\Omega]=0$ (from \eq{2.18}),
\begin{eqnarray}
\Tr(\Omega X_0X) &=&
\Tr((\Omega X)_0X)= -\Tr(\Omega XX_0)=
-\Tr(X_0\Omega X)=
\Tr(-\Omega X_0 X -[X_0,\Omega]X)
\nonumber \\
&=&
-\frac{1}{2}\beta \Tr(\Omega X_{00} X) + O(D^3)
=
\frac{1}{2}\beta \Tr(\Omega X_0^2 ) + O(D^3)
.
\end{eqnarray}
So this is term is of second order in the derivative expansion.

\subsection{Symbols at zero temperature}

A convenient technique to compute the diagonal matrix elements of a
pseudo-differential operator, $\langle x| f(D,X)|x\rangle$, is the method of
symbols \cite{Nepomechie:1984wt,Salcedo:1996qy}.

Let us discuss the zero temperature case first. The Euclidean space-time is
$\R^{\dims}\times\R$. We introduce a momentum basis $|p)$,
\begin{equation}
\langle x|p)= e^{i px},
\qquad
(p|p^\prime)= (2\pi)^{\dimst}\delta(p-p^\prime)
,
\qquad |p)= e^{ip\hat{x}}|0)
,
\end{equation}
and the method of symbols goes as follows
\begin{eqnarray}
\langle x| f(D,X)|x\rangle
&=&
\int \frac{d^{\dimst}p}{(2\pi)^{\dimst}}\,e^{-ipx}
\langle x| f(D,X) |p)
=
\int \frac{d^{\dimst}p}{(2\pi)^{\dimst}}
\langle x| e^{-ip\hat{x}} f(D,X)
e^{ip\hat{x}}|0)
\nonumber\\
&=&
\int \frac{d^{\dimst}p}{(2\pi)^{\dimst}}
\langle x|f(D+ip,X)|0)
.
\label{eq:2.7}
\end{eqnarray}
We have used the relations $e^{-ip\hat{x}}D_\mu e^{ip \hat{x}}= D_\mu+ip_\mu$
and $e^{-ip\hat{x}} X e^{ip\hat{x}}= X$ (because $X$ is multiplicative, i.e.,
it contains no derivatives) and the fact that the map $Y\to e^{-ip\hat{x}} Y
e^{ip\hat{x}}$ is a similarity transformation. In \eq{2.7} $|0)$ is the state
with wavefunction equal to unity, $\langle x|0)=1$.

Due to the property, $\partial_\mu|0)=0$, the quantity $\langle
x|f(D+ip,X)|0)$ is just the {\em symbol} of the pseudo-differential operator
$f(D,X)$ \cite{Salcedo:1996qy}. A very important point is that the operator
$\displaystyle \int \frac{d^{\dimst}p}{(2\pi)^{\dimst}} f(D+ip,X) $ contains
$D_\mu$ only in the form $[D_\mu,~]$. As a consequence, this operator is
automatically gauge covariant and also multiplicative with respect to $x$. As
said, a multiplicative operator is equivalent to a function of
$x$. Specifically, $\langle x|f(\hat{x})|0)= f(x)\langle x|0)= f(x)$.  So
$\langle x|~|0)$ can be left implicit in \eq{2.7}, and one can write just
\begin{equation}
\langle x| f(D,X)|x\rangle
=
\int \frac{d^{\dimst}p}{(2\pi)^{\dimst}} f(D+ip,X)
.
\label{eq:2.8}
\end{equation}
The variable $p_\mu$ represents the momentum carried by the quantum field
$\phi$ running in the loop.

To obtain a covariant derivative expansion, one simply expands the right hand
side of \eq{2.8} in powers of $D_\mu$. Due to gauge invariance, it is
guaranteed that if all $D_\mu$ are brought (e.g.) to the right using $D_\mu
Y=[D_\mu,Y]+Y D_\mu$, at the end all terms with $D_\mu$ not in the form
$[D_\mu,~]$ must vanish after momentum integration. So gauge invariance of the
final result will hold but it is not manifest without momentum integration.

\subsection{Covariant symbols at zero temperature}

The matrix element $\langle x| f(D,X)|x\rangle$ is a gauge covariant quantity,
and its covariant derivative expansion can be obtained by expansion in powers
of $D_\mu$ in \eq{2.8}. However, gauge covariance of the right hand side holds
only after momentum integration: the symbol itself is not covariant. Pletnev
and Banin devised a method to transform the symbol into a covariant one
\cite{Pletnev:1998yu,Salcedo:2006pv}. This is as follows
\begin{equation}
\langle x| f(D,X)|x\rangle
=
\int \frac{d^{\dimst}p}{(2\pi)^{\dimst}} f(\bar{D},\bar{X})
.
\label{eq:2.25}
\end{equation}
with the {\em covariant symbol}
\begin{equation}
f(\bar{D},\bar{X})
= e^{i\partial^p D}e^{-ipx} f(D,X) e^{ipx} e^{-i\partial^p D}
,
\qquad
\partial^p_\mu =\frac{\partial}{\partial p_\mu}
,
\qquad
\partial^p D = D \partial^p = D_\mu \partial^p_\mu 
.
\end{equation}
That is, a further similarity transformation is applied which changes nothing:
the new factor $e^{-i\partial^p D}$ is equivalent to $1$ since no $p_\mu$
lies at its right, and on the other hand the new factor $e^{i\partial^p D}$ is
also equivalent to $1$ by integration by parts. Being a similarity
transformation it can be applied to each block in $f$, i.e.,
$D_\mu\to\bar{D}_\mu$ and $X\to\bar{X}$.
\begin{eqnarray}
\bar{D}_\mu &=&
e^{i\partial^p D}e^{-ipx} D_\mu e^{ipx} e^{-i\partial^p D}
=
e^{i\partial^p D} (D_\mu + ip_\mu) e^{-i\partial^p D}
,
\nonumber \\
\bar{X} &=& 
e^{i\partial^p D}e^{-ipx} X e^{ipx} e^{-i\partial^p D}
=
e^{i\partial^p D} X  e^{-i\partial^p D}
.
\end{eqnarray}

These new operators are directly gauge covariant and multiplicative (with
respect to $x$) without momentum integration. Using a derivative expansion,
they read:
\begin{eqnarray}
\bar{D}_\mu &=& ip_\mu 
+ \sum_{n=1}^\infty
\frac{n}{(n+1)!} i^n F_{\alpha_1\ldots\alpha_n\mu}
\,\partial^p_{\alpha_1}\cdots \partial^p_{\alpha_n}
,
\nonumber \\
\bar{X} &=& \sum_{n=0}^\infty
\frac{1}{n!} i^n X_{\alpha_1\ldots\alpha_n}
\partial^p_{\alpha_1}\cdots \partial^p_{\alpha_n}
.
\label{eq:2.26}
\end{eqnarray}
As can be seen, the covariant symbol is closely related to the Fock-Schwinger
gauge approach. The map $Y\to\bar{Y}$ is an algebra homomorphism that applies
pseudo-differential operators into operators which are covariant and
multiplicative (with respect to $x$). They are derivative operators with
respect to $p_\mu$. Let us stress that, in applications of \eq{2.25}, a
constant function equal to 1 is understood at the right, so that
$\partial^p_\mu\,1=0$.\footnote{Actually, the quantity
  $f^\prime(x,p)=f(\bar{D},\bar{X})1$ is what enters in the computation of
  $\langle x| f(D,X)|x\rangle$. This is an ordinary function of $x$ and $p$
  and so closer to the ordinary symbols, except that it is covariant.}

\subsection{Symbols at finite temperature}

Let us now turn to the finite temperature case. For ordinary symbols one can
proceed as before by introducing a momentum space basis $|p)= |p_0,\vp)$,
where the zeroth component takes values on the Matsubara frequencies: $p_0 =
2\pi n T$ in the bosonic case, $p_0 = (2n+1) \pi T $ in the fermionic case,
with $n\in\Z$. Thus
\begin{equation}
\langle x|p)= e^{i px},
\qquad
(p|p^\prime) = \beta\delta_{p_0,p_0^\prime} (2\pi)^{\dims}\delta(\vp-\vp^\prime)
,
\qquad
|p)= e^{ip\hat{x}}|0)
.
\end{equation}
The method of symbols works as before with the result
\begin{eqnarray}
\langle x| f(D,X)|x\rangle
&=&
T\sum_{p_0}\int \frac{d^{\dims}p}{(2\pi)^{\dims}}
\,
\langle x|f(D+ip,X)|0)
.
\label{eq:2.31}
\end{eqnarray}

Let us remark that $|0)$ is the state $\langle x|0)=1$, regardless of whether
the quantum field in the loop is bosonic or fermionic. The statistics of the
quantum field is contained in the Matsubara frequencies $p_0$. Once again the
operator $T\sum_{p_0}\int \frac{d^{\dims}p}{(2\pi)^{\dims}} f(D+ip,X)$ is
actually multiplicative and $\langle x|~|0)$ can be omitted
\begin{eqnarray}
\langle x| f(D,X)|x\rangle
&=&
T\sum_{p_0}\int \frac{d^{\dims}p}{(2\pi)^{\dims}} \, f(D+ip,X)
.
\label{eq:2.32}
\end{eqnarray}
Also $\langle x| f(D,X)|x\rangle$ is still gauge covariant.

In previous works we have discussed the effect of the finite temperature,
i.e., the replacement of an integral over $p_0$ on $\R$ to a sum of $p_0$ over
Matsubara frequencies. As in the zero temperature case, after integration over
$\vp$, the operator $\vD$ appears only in the form $[\vD,~]$. The reason is
that obviously if one replaces $\vD$ by $\vD+i\va$, $\va$ being a constant
$c$-number, the replacement has no effect owing to the integration over $\vp$
on $\R^{\dims}$. However, the same argument fails for $D_0$ (the zeroth
component of the gauge covariant derivative) since $p_0$ is a discrete
variable at finite temperature. Still, due to the sum over the Matsubara
frequencies, the expression must be periodic in the variable $D_0$ with period
$2\pi i T$. This not only permits a dependence on $[D_0,~]$ but also on
$e^{-\beta D_0}=\Omega$, i.e., on the Polyakov loop.

Let us discuss how to use the ordinary symbols to obtain the diagonal matrix
elements at finite temperature \cite{GarciaRecio:2000gt}. The main issue is
the gauge invariance. In the method of symbols, \eq{2.32}, gauge invariance of
$\langle x| f(D,X)|x\rangle$ is manifest only after the integral on $\vp$ and
the sum on $p_0$ are carried out. In $f(D+ip,X)$, $\vD$ can be dealt with as
in the zero temperature case to yield $[\vD,~]$ after integration on
$\vp$. This produces an expression of the type $f_1(D_0+ip_0,[\vD,~],X)$.  As
described in \cite{GarciaRecio:2000gt,Megias:2003ui}, a method suitable to deal with $D_0$ to
obtain a derivative expansion, is to move $D_0$ to the left (using the
identity $YD_0=D_0Y-[D_0,Y]$). In this way one ends up with expressions of the
type $f_2(D_0+ip_0;[D_0,~],[\vD,~],X)$ where $D_0+ip_0$ is only at the left,
rather than all over the expression.  Summing now over the Matsubara
frequencies produces a dependence on $e^{-\beta D_0}=\Omega$ and finally a
covariant expression of the type $f_3(\Omega,[D_\mu,~],X)$ with all
$\Omega(x)$ at the left. This is the form in \eq{2.7b} or in \eq{2.7c}.

\subsection{Covariant symbols at finite temperature}
\label{subsec:II.E}

The method just described at the end of the previous section is rather
cumbersome, so a method of covariant symbols at finite temperature would be
advisable, namely, a method providing manifestly multiplicative and gauge
invariant terms.  The problem is that the method of Pletnev and Banin does not
directly apply at finite temperature, since $p_0$ is a discrete variable and
$\partial^p_0$ is not defined. Presently we show how to extend the method to
the finite temperature case.

One idea is to change the sum over Matsubara frequencies by appropriate
integrals on the complex plane \cite{Landsman:1986uw}. In this way the
derivative with respect to $p_0$ is defined. This method works but we can
obtain the final result in a simpler manner.

Let $\omega_n$ be the Matsubara frequencies, bosonic ($\omega_n=2 n\pi T$) or
fermionic ($w_n= (2n+1)\pi T$). Then let
\begin{equation}
h_M(p_0) = \sum_n 2\pi T\delta(p_0-\omega_n)
\label{eq:2.33}
.
\end{equation}
(There is a bosonic version and a fermionic version of this function.) 

Using the function $h_M$ we can write \eq{2.31} as
\begin{eqnarray}
\langle x| f(D,X)|x\rangle
&=&
\int \frac{d^{\dimst}p}{(2\pi)^{\dimst}}
\,
h_M(p_0)
\langle x|f(D+ip,X)|0)
.
\end{eqnarray}
Now we can proceed to make a further similarity transformation, as at zero
temperature (and valid by the same reasons)
\begin{eqnarray}
\langle x| f(D,X)|x\rangle
&=&
\int \frac{d^{\dimst}p}{(2\pi)^{\dimst}}
\,
\langle x|
e^{i\partial^p D}
h_M(p_0)
f(D+ip,X)
e^{-i\partial^p D}
|0)
\nonumber \\
&=&
\int \frac{d^{\dimst}p}{(2\pi)^{\dimst}}
\,
\langle x|
e^{i\partial^p D}
h_M(p_0)
e^{-i\partial^p D}
f(\bar{D},\bar{X})
|0)
.
\end{eqnarray}

This can be simplified by working out the $h_M(p_0)$ term:
\begin{equation}
e^{i\partial^p D}
p_0
e^{-i\partial^p D}
= p_0 + iD_0 -\frac{1}{2}iF_{0i}\partial^p_i
+ \frac{1}{6}F_{\mu 0i} \partial^p_\mu \partial^p_i + \cdots
\end{equation}
hence
\begin{equation}
e^{i\partial^p D}
h_M(p_0)
e^{-i\partial^p D}
= h_M(p_0 + iD_0) + O(\partial^p_i)
.
\end{equation}
The point is that, due to the integration on $\vp$, all $\partial_i^p$ at the
left (no $\vp$ lies at the left of the $\partial^p_i$) can be set to zero, and
so
\begin{eqnarray}
\langle x| f(D,X)|x\rangle
&=&
\int \frac{d^{\dimst}p}{(2\pi)^{\dimst}}
\,
\langle x|
h_M(p_0 + iD_0)
f(\bar{D},\bar{X})
|0)
.
\label{eq:2.36}
\end{eqnarray}

The expression \eq{2.36} is of great interest. $\bar{D}$ and $\bar{X}$ are the
same covariant symbols as at zero temperature, and so they are {\em Lorentz}
covariant (if the original pseudo-differential operator $\hat{f}$ is).  They
are also multiplicative with respect to $x$-space and manifestly gauge
covariant. On the other hand the $D_0$ dependence at the left is also
multiplicative: under the shift $D_0 \to D_0 + 2\pi i n T$ the expression is
unchanged due to periodicity of $h_M$ (even without integral over
$p_0$). Therefore, the dependence is really on the periodic variable
$e^{-\beta D_0}=\Omega$. That is, one can also write\footnote{Once again, in
  \eq{2.38}, a constant function equal to 1 is implicit at the right, so that
  $\partial^p_\mu\,1=0$.}
\begin{equation}
\langle x| f(D,X)|x\rangle
=
\int \frac{d^{\dimst}p}{(2\pi)^{\dimst}}
\,
h_M(p_0 -iT\log\Omega)
\,
f(\bar{D},\bar{X})
.
\label{eq:2.38}
\end{equation}

This expression is already of the form required, gauge covariant and with
$\Omega$ at the left, suitable to take the expansions in \eq{2.7b} or
\eq{2.7c}.

For convenience let us introduce the auxiliary multiplicative operator (a
matrix in internal space)
\begin{equation}
Q(x) = iT\log\Omega(x)
\label{eq:2.39}
.
\end{equation}
This is many-valued but in practice it appears in periodic functions so that
the result is always a one-valued function of $\Omega$. $Q$ is Hermitian, up
to many-valuation, $\Omega$ being unitary. \Eq{2.38} takes the form
\begin{equation}
\langle x| f(D,X)|x\rangle
=
\int \frac{d^{\dimst}p}{(2\pi)^{\dimst}}
\,
h_M(p_0 -Q)
\,
f(\bar{D},\bar{X})
.
\label{eq:2.38a}
\end{equation}

It is possible to define also the quantity $Q_0$ as the operator $Q$ {\em
  placed at the left of all other operators}, that is, labeled to indicate
``at position zero''. There can be no confusion with our previous convention
of a label 0 indicating a temporal covariant derivative since $[D_0,Q]=0$ due
to $[D_0,\Omega]=0$.  The point is that {\em $Q_0$ is a c-number}: it can be
put in any order in an expression with the same result. Hence we can shift the
variable $p_0$ by an amount $Q_0$. This allows us to write
\begin{eqnarray}
\langle x| f(D,X)|x\rangle
&=&
T\sum_{p_0}\int \frac{d^{\dims}p}{(2\pi)^{\dims}}
\,
f(\bar{D},\bar{X})\big|_{p_0 \to p_0+Q_0}
\nonumber \\
&=&
T\sum_{p_0}\int \frac{d^{\dims}p}{(2\pi)^{\dims}}
\,
f(\bar{D}_0 + i Q_0,\bar{D}_i,\bar{X})
.
\label{eq:2.44}
\end{eqnarray}
(In the last equality we have used that the variable $p_0$ does not appear in
$\bar{D}_0 - i p_0$, $\bar{D}_i$ or $\bar{X}$.)

In \eq{2.38a} one can carry out the momentum derivatives $\partial^p_\mu$
implied by $\bar{D}_\mu$ and $\bar{X}$. The derivatives $\partial^p_i$ can be
taken to the right or to the left, by parts. The temporal derivative
$\partial^p_0$ can only be taken to the right, if the form of $h_M(p_0-D_0)$
is to be preserved. Taking all of the $\partial^p_\mu$ to the right has the
virtue of leaving an ordinary function $f^\prime(x,p)$ which is {\em
  temperature independent, and manifestly Lorentz and gauge covariant}:
\begin{eqnarray}
\langle x| f(D,X)|x\rangle
&=&
\int \frac{d^{\dimst}p}{(2\pi)^{\dimst}}
\,
h_M(p_0-Q)
\,
f^\prime(x,p)
.
\label{eq:2.38b}
\end{eqnarray}

The eqs. (\ref{eq:2.38a}) or (\ref{eq:2.38b}) solve the problem of using gauge
covariant symbols at finite temperature. In addition the breaking of Lorentz
covariance is minimal. The zero temperature limit is recovered by setting
$h_M$ to unity.

\subsection{Polyakov loop and real time thermal field theory}
\label{subsec:II.F}

Mathematically, the imaginary time formalism is the simplest approach to
quantum field theory at finite temperature. The real time approach (in its
various versions) is more involved but better suited for time dependent
observables \cite{Landsman:1986uw}.

In that approach the frequency is a continuous variable, rather than
discrete. In the expressions derived for the covariant symbols at finite
temperature the sum over Matsubara frequencies can be traded by integrals
using well known relations (eqs.~(2.3.22-24) of
\cite{Landsman:1986uw}). Starting from \eq{2.44}, we find
\begin{eqnarray}
\langle x| f(D,X)|x\rangle
&=&
\int \frac{d^{\dims}p}{(2\pi)^{\dims}}
\Big(
\int\frac{dp_0}{2\pi}
+
\int_{C_+}\frac{dp_0}{2\pi}n(-p_0)
-
\int_{C_-}\frac{dp_0}{2\pi}n(p_0)
\Big)
f(\bar{D}_0 + i Q_0,\bar{D}_i,\bar{X})
.
\end{eqnarray}
Here
\begin{equation}
n(p_0) = \frac{1}{\pm e^{i\beta p_0} -1}
,
\end{equation}
($\pm$ for bosons or fermions, respectively). The first frequency integral is
along the $p_0$ real axis, whereas the contours $C_\pm$ enclose only the
singularities of $f$ as a function of $p_0$, in the half planes ${\rm Im\,}
p_0>0$ and ${\rm Im\,} p_0<0$, respectively. Undoing the shift $p_0\to
p_0+Q_0$, gives
\begin{eqnarray}
\langle x| f(D,X)|x\rangle
&=&
\int \frac{d^{\dims}p}{(2\pi)^{\dims}}
\Big(
\int\frac{dp_0}{2\pi}
+
\int_{C_+}\frac{dp_0}{2\pi}
\frac{1}{\pm \Omega^{-1} e^{-i\beta p_0}-1}
-
\int_{C_-}\frac{dp_0}{2\pi}
\frac{1}{\pm \Omega e^{i\beta p_0}-1}
\Big)
f(\bar{D},\bar{X})
.
\label{eq:2.48}
\end{eqnarray}
($C_\pm$ are as before for the new $f$, since $Q_0$ is real.)

This expression is not yet in the form of the real time formalism but it is
closer to it. Upon Wick rotation, factors of the type $n(p_0)$ should appear
there in the propagators through the thermal occupation numbers, while the
integral over the real axis should come from the zero temperature part of the
thermal propagators \cite{Landsman:1986uw}.

The connection with the real time formalism is, of course, of great interest,
and worth studying.  We do not pursue this subject any further in the present
work, but in view of \eq{2.48} one can conjecture that the fields in the form
$f(\bar{D},\bar{X})$, including time covariant derivatives, will follow the
pattern of ordinary local external fields as treated in the real time
formalism. On the other hand, the occupation number will pick up a Polyakov
loop following the prescription $e^{\beta p_0} \to \Omega e^{\beta p_0}$. This
automatically produces the correct coupling of the chemical potential,
$e^{\beta\mu}$, by means of the prescription $A_0(\vx) \to A_0(\vx)-\mu$
(since $-A_0$ essentially represents a local and possibly non Abelian chemical
potential).

The meaning of $\Omega(\vx)$ in the real time context needs to be
elucidated. If the configuration of the external fields is stationary,
essentially the imaginary time formula gives already the result. In this case
$\Omega(\vx)= e^{-\beta A_0(\vx)}$ with $A_0(\vx)$ Hermitian. $A_0$ is the
same variable in the Euclidean and Minkowski versions, however, in the
Minkowski case this variable is taken along the real axis of its complex
plane, while in the Euclidean version it is preferable to work with the
variable extended to the imaginary axis. The imaginary time integration in the
definition of $\Omega$ is not rotated to real time because it comes from the
factor $e^{-\beta H}$ in the partition function and so from an evolution in
imaginary time from $t_0$ to $t_0-i\beta$ (Kubo-Martin-Schwinger condition
\cite{Kubo:1957mj}). Therefore takes the same form any finite temperature
formulation.

Let us consider now the more general case of non stationary configurations.
In the closed-path approach \cite{Schwinger:1960qe,Keldysh:1964ud} one starts
with a thermal mixed state at time $t=t_0$. This implies that the system is
stationary for $t < t_0$. Measurements are taken at later times, where also
time dependent sources may act. By assumption of thermal equilibrium, either
the Hamiltonian is stationary for $t<t_0$, or it is so upon taken a suitable
gauge transformation. Then, for $t<t_0$, $A_0(\vx)$ is well defined modulo
stationary gauge transformations and this defines $\Omega(\vx)= e^{-\beta
  A_0(\vx)}$ which also transforms covariantly. Further gauge transformations,
for instance carrying $A_0$ to zero, exist but they are not stationary and so
they would introduce a time dependence in the other components of the gauge
connection (and possible on other gauge covariant fields). Therefore such
transformations are not allowed for $t<t_0$ and $\Omega$ is well
defined. Because the external fields configuration is not required to be
stationary for $t>t_0$ one can choose a continuous gauge in which $A_0(\vx)$
takes the same value at all times. (Because any $A_0(\vx,t)$ can be brought
to zero by means of a suitable gauge transformation, any configuration
$A_0(\vx,t)$ can be transformed into any other.) This shows that $\Omega(\vx)$
is also present in the real time approach and this is the quantity that will
appear with $e^{\beta p_0}$ in the propagators.

\section{Heat kernel at finite temperature}
\label{sec:3}

\subsection{Diagonal coefficients}

\subsubsection{Expansions of the heat kernel}

Let $K=D^2+X$ be the Klein-Gordon operator as in \eq{2.1}. The heat kernel is
the solution of the associated heat equation $\partial_\tau G(\tau) = K
G(\tau)$, $G(0)=1$, $\tau\ge 0$, with solution $G(\tau)= \exp(\tau
K)$.\footnote{For the Klein-Gordon operator the parameter $\tau$ has
  dimensions of inverse mass squared, nevertheless it is called the
  Fock-Schwinger proper time \cite{Itzykson:1980bk} since in the heat kernel
  equation it plays the role a time with corresponding Hamiltonian $iK$ acting
  in the Hilbert space spanned by $|x\rangle$.} From the heat kernel one can
recover the propagator, $K^{-1}$, and the effective action, $\Tr\log K$.

The diagonal matrix elements of the heat kernel (at zero or at finite
temperature) can be expanded classifying the terms by their mass dimension:
\begin{equation}
\langle x|e^{\tau K}|x\rangle
 =
\frac{1}{(4\pi\tau)^{\dimst/2}} \sum_n \tau^n a_n(x;\tau) 
.
\end{equation}
Each $a_n$ has dimension $2n$ and depends on the temperature. The expansion is
asymptotic. At zero temperature this is equivalent to an expansion in powers
of $\tau$, and this is just the standard heat-kernel expansion. In general the
$a_n$ depend also on $\tau$ and $T$. The order of the term is defined by the
mass dimension carried by the external fields. Hence by dimensional counting,
the coefficient can only depend on the combination $\tau T^2$. A remarkable
property of the heat kernel coefficients is that they do not depend explicitly
on the space-time dimension. This property is preserved at finite temperature.

At zero temperature the index $n$ takes non negative integer
values.\footnote{There are half integer orders in the presence of
  boundaries. We only consider boundaryless manifolds throughout.}  However,
at finite temperature $n$ can also take (positive) half-integer values. This
follows from breaking of Lorentz invariance down to rotational invariance; at
finite temperature an odd number of time derivatives is not forbidden. The
expansion at finite temperature has been computed in
\cite{Megias:2002vr,Megias:2003ui} through dimension 6. So for
instance,\footnote{Regarding conventions, let us note that what is called here
  $K$ and $X$ corresponds to $-K$ and $-M$ in
  \cite{Megias:2002vr,Megias:2003ui}.  The functions $\xi_n$ are similar to
  the $\varphi_n$ in \cite{Megias:2002vr,Megias:2003ui} except that they
  involve the Hermite polynomials.}
\begin{eqnarray}
a_0 &=& \xi_0,
\nonumber \\
a_{1/2} &=& 0,
\nonumber \\
a_1 &=&  \xi_0 X,
\nonumber \\
a_{3/2} &=&  \frac{1}{2}\xi_1( X_0 + E_{ii})
.
\end{eqnarray}

The electric field, $E_i(x)$, is defined as $F_{0i}(x)$, hence
$E_{ii}=-F_{ii0}$. On the other hand, the $\xi_n$ are dimensionless functions
of the Polyakov loop defined as sums over the (bosonic or fermionic) Matsubara
frequencies:
\begin{eqnarray}
  \xi_n &=& (4\pi\tau)^{1/2} (-i)^n 2^{-n/2} T\sum_{p_0}  
H_n(\sqrt{2\tau}(p_0+Q))
  e^{-\tau(p_0+Q)^2}
\nonumber
\\
&=&
2^{-n/2}\sum_{k\in\Z}H_n(k/\sqrt{2\tau T^2}) e^{-k^2/(4\tau T^2)}(\pm\Omega)^k
,
\qquad
n=0,1,2,\ldots
\label{eq:3.3}
\end{eqnarray}
$Q$ was introduced in \eq{2.39}. $H_n$ refers to the $n$-th Hermite polynomial
(with normalization $H_1(x)=2x$). The $\pm$ refers to bosonic or fermionic
case, respectively. The two forms of $\xi_n$ in \eq{3.3} are related by
Poisson summation formula.  The $\xi_n$ are one-valued functions of $\Omega$
and of $\tau T^2$.  They are real (Hermitian) for even $n$ and imaginary
(antiHermitian) for odd $n$. In addition, they are even or odd under
$\Omega\to \Omega^{-1}$ for even or odd $n$, respectively.  In the zero
temperature limit
\begin{equation}
\xi_n^{T=0} = 2^{-n/2} H_n(0)
,
\end{equation}
so odd orders vanish in this limit.

It will be also convenient to define the following auxiliary combinations
\begin{equation}
\bar{\xi}_1  = \xi_1
,\quad
\bar{\xi}_2  = \xi_2+\xi_0
,\quad
\bar{\xi}_3  = \xi_3+3\xi_1
,\quad
\bar{\xi}_4  = \xi_4+6\xi_2+3\xi_0
.
\end{equation}
They vanish at zero temperature. (However, the $\bar{\xi}_n$ do not vanish at
finite temperature for $\Omega=1$ for even orders.)

The {\em derivative expansion} of the heat kernel (at zero or finite
temperature) takes the form
\begin{equation}
\langle x|e^{\tau K}|x\rangle
 =
\frac{1}{(4\pi\tau)^{\dimst/2}} \sum_n \tau^n A_n(x;\tau)
,
\label{eq:3.6}
\end{equation}
where the coefficient $A_n$ contains $2n$ derivatives, as well as the
Polyakov loop (placed at the left) and any number of $X$. By dimensional
counting, besides the derivatives, $A_n(x;\tau)$ depends on $\tau X$ and $\tau
T^2$ and $\Omega$.  This is an asymptotic expansion. Once again, at zero
temperature the index $n$ takes only nonnegative integer values, whereas at
finite temperature half-integer values are allowed. The derivative expansion
coefficients $A_n$ are also independent of the space-time dimension, at zero
or finite temperature.

The expansion at zero temperature has been considered in \cite{Salcedo:2004yh}
to four derivatives (and six derivatives for the traced coefficients). For
instance,
\begin{eqnarray}
A_0 &=& I_1,
\nonumber \\
A_1 &=& \tau I_{2,2} \, X_{\mu\mu} + 2 \tau^2 I_{2,1,2} \, X_\mu^2 
.
\end{eqnarray}

The coefficients $I_1$, $I_{2,2}$ and $I_{2,1,2}$ are functions of the labeled
operators $X_1$ in the first case, $X_1$, $X_2$ in the second, and $X_1$,
$X_2$ and $X_3$ in third case. In general, these coefficients are defined as
follows \cite{Salcedo:2004yh}
\begin{equation}
I_{r_1,r_2,\ldots,r_n} = 
\int_\Gamma\frac{dz}{2\pi i} e^z N_1^{r_1}N_2^{r_2}\cdots N_n^{r_n}
,
\qquad
r_i=0,1,2,\ldots
\label{eq:3.7}
\end{equation}
where
\begin{equation}
N_i = (z - \tau X_i)^{-1}
,
\end{equation}
and $\Gamma$ is a positively oriented simple closed path enclosing all the
$X_i$.\footnote{This $\Gamma$ is not to be confused with the effective action
  functional introduced in \eq{2.3}.} Explicitly
\begin{equation}
I_{r_1,r_2,\ldots,r_n} = 
\tau^{1 - \sum_{i=1}^nr_i}
\sum_{i=1}^n\frac{1}{(r_i-1)!}\frac{d^{r_i-1}}{dX_i^{r_i-1}}\frac{e^{\tau X_i}}
{\prod_{j\not=i}(X_i-X_j)^{r_j}}
.
\end{equation}
The functions $I_{r_1,r_2,\ldots,r_n}$ are analytical on the $X_i$ even at
coincident points (as follows from \eq{3.7}, the singularities at $X_i=X_j$
are removable) and satisfy recurrence relations. Instances at lower orders are
\begin{eqnarray}
I_r &=& \frac{e^{\tau X_1}}{(r-1)!},
\qquad
r= 0,1,2,\ldots
\nonumber \\
I_{2,2} &=& 
\frac{1}{\tau^2}\frac{e^{\tau X_1} + e^{\tau X_2}}{(X_1-X_2)^2}
-
\frac{2}{\tau^3}\frac{e^{\tau X_1} - e^{\tau X_2}}{(X_1-X_2)^3}
.
\end{eqnarray}

\subsubsection{Derivative expansion at finite temperature}
\label{subsec:IIIA}

The coefficients $A_n$ at finite temperature are not yet known. They can be
computed from scratch by using the tools previously described. To this end we
use an integral representation of the heat kernel
\begin{equation}
e^{\tau K} = \int_\Gamma \frac{dz}{2\pi i} \frac{e^{\tau z}}{z-D^2-X}
,
\end{equation}
where the path $\Gamma$ is positively oriented and encloses the eigenvalues of
$K$ (the concrete realization of this requirement will be clear below).

Applying the method developed in section \ref{subsec:II.E} for covariant
symbols at finite temperature, and in particular \eq{2.38a}, we can write
\begin{equation}
\langle x|e^{\tau K}|x\rangle 
=
\int_\Gamma \frac{dz}{2\pi i}
\,
\int \frac{d^{\dimst}p}{(2\pi)^{\dimst}}h_M(p_0-Q)
\frac{e^{\tau z}}{z-\bar{D}^2-\bar{X}}
.
\label{eq:3.13}
\end{equation}
\ignore{
$Q_0$ means that the operator $Q$ is to be put at the left of all other
operators.}

Using the explicit expressions of the covariant symbols of $\D_\mu$ and $X$ in
\eq{2.26}, it is simple to carry out an expansion with terms classified by the
number of covariant derivatives they have (regardless of the number of $X$ or
$Q$). Specifically,\footnote{Alternatively one can use the formulas of
  Appendix \ref{app:C} for the covariant symbols of $K$ and $(z-K)^{-1}$.}
removing the zeroth order contributions in $\bar{D}_\mu$ and $\bar{X}$,
\begin{equation}
\bar{D}^\prime_\mu = \bar{D}_\mu -ip_\mu = O(D^2)
,
\qquad
\bar{X}^\prime = \bar{X} - X = O(D)
,
\end{equation}
we can write
\begin{eqnarray}
(z-\bar{D}^2-\bar{X})^{-1}
&=&
\left(
N^{-1}
-i\{p_\mu , \bar{D}^\prime_\mu\} - \bar{D}^{\prime 2}
- \bar{X}^\prime
\right)^{-1}
\nonumber\\
&=&
\sum_{n=0}^\infty
N((i\{p_\mu , \bar{D}^\prime_\mu\} + \bar{D}^{\prime 2}
+ \bar{X}^\prime)N)^n
,
\label{eq:3.15}
\end{eqnarray}
where we have introduced the quantity
\begin{equation}
N = (z + p^2 -X)^{-1}
.
\end{equation}

Let us spell out the details for $A_{1/2}$ (i.e., one derivative). Picking up
the terms with precisely one derivative in \eq{3.13} gives (using \eq{3.15}
and \eq{2.26})
\begin{equation}
\langle x|e^{\tau K}|x\rangle_{1/2}
=
\int_\Gamma \frac{dz}{2\pi i}
\,
\int \frac{d^{\dimst}p}{(2\pi)^{\dimst}}h_M(p_0-Q)
e^{\tau z}
N
i X_\mu \partial^p_\mu
N
.
\end{equation}
Further, applying the identity
\begin{equation}
(\partial^p_\mu N) = -2p_\mu N^2
,
\end{equation}
yields
\begin{equation}
\langle x|e^{\tau K}|x\rangle_{1/2}
=
\int_\Gamma \frac{dz}{2\pi i}
\,
\int \frac{d^{\dimst}p}{(2\pi)^{\dimst}}h_M(p_0-Q)
e^{\tau z}
(-2i) p_\mu N X_\mu N^2
.
\end{equation}

Next, let us apply the shift $z\to z-p^2$, so that
\begin{equation}
\langle x|e^{\tau K}|x\rangle_{1/2}
=
\int \frac{d^{\dimst}p}{(2\pi)^{\dimst}}h_M(p_0-Q)
e^{-\tau p^2} (-2i) p_\mu
\int_\Gamma \frac{dz}{2\pi i}
\,
e^{\tau z}
 N X_\mu N^2
,
\end{equation}
where
\begin{equation}
N = (z -X)^{-1}
.
\end{equation}

Now the $z$ and $p$ integrals are independent. For the $z$ integral, the
definition of $I_{r_1,\ldots,r_n}$ in \eq{3.7} applies:
\begin{equation}
\int_\Gamma \frac{dz}{2\pi i}
\,
e^{\tau z}
 N X_\mu N^2
=
\tau^2 I_{1,2} X_\mu
.
\end{equation}

For the $p$ integral,
the definition of $\xi_n$ in \eq{3.3} applies:
\begin{eqnarray}
\int \frac{d^{\dimst}p}{(2\pi)^{\dimst}}h_M(p_0-Q)
e^{-\tau p^2} (-2i) p_\mu
&=&
(-2i)\delta_{\mu 0}
\frac{1}{(4\pi\tau)^{(\dims)/2}}
\int \frac{d p_0}{2\pi}h_M(p_0-Q)
e^{-\tau p_0^2} p_0 
\nonumber\\
&=&
\delta_{\mu 0}
\frac{1}{(4\pi\tau)^{\dimst/2}} 
\tau^{-1/2}\xi_1
.
\end{eqnarray}
Therefore,
\begin{equation}
\langle x|e^{\tau K}|x\rangle_{1/2}
=
\frac{1}{(4\pi\tau)^{\dimst/2}}
\tau^{3/2}
\xi_1 I_{1,2} X_0
.
\end{equation}
or according to \eq{3.6},
\begin{equation}
A_{1/2}
=
\tau
\xi_1 I_{1,2} X_0
.
\end{equation}

In what follows we use units $\tau=1$. $\tau$ can be easily restored by
dimensional considerations.

Using the method just described and the formulas in Appendix \ref{app:B} for
the momentum integrals, we find to three derivatives
\begin{eqnarray}
A_0 &=& I_1 \,\xi_0
,
\nonumber \\
A_{1/2} &=& I_{1,2}\bar{\xi}_1\, X_0
,
\nonumber \\
A_1 &=& 
I_{2,2}\, \xi_0\,X_{\mu\mu}
+
2 I_{2,1,2}\, \xi_0\,X_{\mu}X_{\mu}
+ I_{1,3}\,\bar{\xi}_2\,X_{00}
+(2 I_{1,1,3} + I_{1,2,2})\,\bar{\xi}_2\,X_0 X_0
,
\nonumber \\
A_{3/2}
&=&
I_{1,1,2} \bar{\xi}_1 F_{0 \mu } X_{\mu }
+\frac{1}{3} I_{1,2} \bar{\xi}_1 F_{\mu  0 \mu }
+\frac{2}{3} I_{2,3} \bar{\xi}_1 (X_{0 \mu  \mu } 
+ X_{\mu  0 \mu } + X_{\mu  \mu  0})
\nonumber \\ &&
+\left(2 I_{1,1,3}-6 I_{1,1,4}
+I_{1,2,2}-2 I_{1,2,3}\right) \bar{\xi}_1 X_0 X_{\mu  \mu }
+2 I_{2,1,3} \bar{\xi}_1 ( X_{\mu } X_{0 \mu} + X_{\mu } X_{\mu  0})
\nonumber \\ &&
+\left(2 I_{2,1,3}+I_{2,2,2}\right) \bar{\xi}_1 
(X_{0 \mu } X_{\mu}+ X_{\mu  0} X_{\mu} + X_{\mu  \mu } X_0)
\nonumber \\ &&
+\left(4 I_{1,2,1,3}+2 I_{1,2,2,2}+4 I_{1,3,1,2}+4I_{2,1,1,3}
  +2 I_{2,1,2,2}+2 I_{2,2,1,2}\right) \bar{\xi}_1 X_0 X_{\mu } X_{\mu}
\nonumber \\ &&
+\left(4 I_{2,1,1,3}+2 I_{2,1,2,2}\right) \bar{\xi}_1 
(X_{\mu } X_0 X_{\mu} + X_{\mu } X_{\mu } X_0)
+I_{1,4} \bar{\xi}_3 X_{0 0 0}
\nonumber \\ &&
+\left(3 I_{1,1,4}+ I_{1,2,3}\right) \bar{\xi}_3 X_0 X_{0 0}
\nonumber \\ &&
+\left(3 I_{1,1,4}+2 I_{1,2,3}+I_{1,3,2}\right) 
 \bar{\xi}_3  X_{0 0} X_0
\nonumber \\ &&
+\left(6 I_{1,1,1,4}+4 I_{1,1,2,3}+2 I_{1,1,3,2}+2I_{1,2,1,3}+I_{1,2,2,2}\right) 
 \bar{\xi}_3 X_0 X_0 X_0
.
\label{eq:3.26}
\end{eqnarray}
Important remark: For notational convenience we have written $\xi_0$ or
$\bar{\xi}_n$ at the right of the $I_{r_1,r_2,\ldots}$, {\em but actually
  these operators are at the left of the expression}. So $A_0=\xi_0 I_1$,
$A_{1/2} = \bar{\xi}_1I_{1,2}X_0$, $A_1= \xi_0 I_{2,2} X_{\mu\mu}+\cdots$,
etc.

We have also computed the term with four derivatives $A_2$, but this term is
too long to be quoted here (about 90 terms). The four derivative term is given
below for the traced heat kernel coefficients.

The heat kernel (and in fact $\langle x|f(K)|x\rangle$ for any $f(z)$) is
symmetric under left-right transposition of operators (or Hermitian if $K$ is
Hermitian and $f(z)$ is real). At zero temperature (putting
$\bar{\xi}_n\to 0$ and $\xi_0\to 1$) the symmetry is manifest. For instance,
the term $I_{2,2}X_{\mu\mu}+2I_{2,1,2}X_\mu X_\mu$ is symmetric.  The symmetry
is not manifest at finite temperature because it is hidden after having chosen
to put the Polyakov loop to the left. In addition, transposition and
subsequent move of the Polyakov loop to the left in a term $A_n$ produces new
terms of higher order. For instance, to first order in the derivative
expansion,
\begin{equation}
(\xi_0I_1+\bar{\xi}_1I_{1,2}X_0)^T-(\xi_0I_1+\bar{\xi}_1I_{1,2}X_0) =
[I_1,\xi_0]-\bar{\xi}_1(I_{1,2}+I_{2,1})X_0+O(D^2).
\end{equation}
From \eq{2.18}, $\displaystyle [I_1,\xi_0]=i\frac{d\xi_0}{dQ}[D_0,I_1]$. Use
of $\displaystyle \frac{d\xi_0}{dQ}=-i\bar{\xi}_1$, ~$[D_0,I_1]=I_{1,1}X_0$,
and $I_{1,2}+I_{2,1}=I_{1,1}$, shows that the symmetry holds to the order
considered.

The $I_{r_1,\ldots,r_n}$ are not linearly independent, so although the
coefficients in $A_n$ are well defined functions of the labeled operators,
$X_i$, their expression in terms of the $I_{r_1,\ldots,r_n}$ is not unique.

The heat kernel does not depend on the prescription adopted regarding the
position of the Polyakov loop (our choice throughout has been to put it at the
left) but the value of each $A_n$ will be different for different
prescriptions.

The Polyakov loop comes out automatically in the expressions, and, as noted in
the Introduction, its presence is required to accommodate the chemical
potential. Nevertheless, it is also a nuisance and so the possibility
suggests itself to dispose of the Polyakov loop dependence just by setting
$\Omega=1$ in the formulas by hand. We call this the quenched
approximation. If this is done, the $\xi_n$ become ordinary functions of the
temperature (rather than operators) for even $n$ and zero for odd
$n$. Unfortunately, the result of quenching will depend on the prescription
adopted (regarding the position of the Polyakov loop) and in particular the
left-right symmetry can be lost. In fact, the expressions are fully consistent
only when the full Polyakov loop dependence is retained. For the traced heat
kernel, and so for the effective action, setting $\Omega$ to unity by hand is
also dangerous. Due to the cyclic property, the same expression can be written
in several equivalent but different ways. Consequently, the result obtained by
setting $\Omega=1$ by hand will yield different results in each case. This
point is further discussed in Section \ref{subsubsec:IIIC4}.

\subsection{Traced heat kernel coefficients}
\label{subsec:IIIB}

It is also of interest to compute the trace of the heat kernel and this
produces shortest expressions. Specifically (remember that we have set
$\tau=1$)

\begin{equation}
\Tr (e^K) = 
\int d^{\dimst}x \,\tr\langle x|e^K|x\rangle
 =
\frac{1}{(4\pi)^{\dimst/2}} \sum_n \int d^{\dimst}x \,\tr B_n(x)
.
\label{eq:3.27}
\end{equation}

The choice $B_n=A_n$ is of course correct, but some simplification in the form
of the coefficients $B_n$ can be achieved by using integration by parts and
the cyclic property of the trace. When using this freedom, the functions
$\xi_n$ should be moved to the left by using the identity in \eq{2.18}.

Note that the $A_n$ can be recovered from the $B_n$ using the identity
\begin{equation}
\langle x|e^K|x\rangle =
\frac{\delta \Tr (e^K)}{\delta X(x)}
\,.
\label{eq:3.29}
\end{equation}
This equality holds separately at each order in the derivative expansion.

For convenience, we separate in $B_n$ terms with a contribution at zero
temperature from those which vanish in that limit,
\begin{equation}
B_n= B^{(0)}_n + B^{(T)}_n. 
\end{equation}
The $B^{(0)}_n$ vanish for half-integer $n$, and are of the form $\xi_0
(B_n\big|_{T=0})$, while $B^{(T)}_n\big|_{T=0}=0$.\footnote{Note that the
  definition $B^{(0)}_n=\xi_0 (B_n\big|_{T=0})$ would be ambiguous, since
  $B_n\big|_{T=0}$ can be written in different ways which are equivalent
  inside the trace, but not in $\xi_0 (B_n\big|_{T=0})$.}

The results are as follows:

\begin{eqnarray}
B^{(0)}_0 &=& I_{1}{\xi}_0 
,
\nonumber \\ 
B^{(0)}_1 &=&
-\frac{1}{2}I_{1,2,1}{\xi}_0 X_\mu X_\mu
,
\nonumber \\ 
B^{(0)}_2 &=&
2 I_{2,2,2,0} \xi_0 X_{\mu } X_{\nu } F_{\mu \nu }
+\frac{1}{2} I_{2,2,0} \xi_0 F_{\mu \nu } F_{\mu \nu }
+I_{3,3,0} \xi_0 X_{\mu \mu } X_{\nu \nu }
\nonumber \\ &&
+4 I_{3,1,3,0} \xi_0 X_{\mu } X_{\mu } X_{\nu \nu }
+\frac{1}{2} I_{2,2,2,2,0} \xi_0 X_{\mu } X_{\nu } X_{\mu } X_{\nu }
\nonumber \\ &&
+ ( 4 I_{3,1,3,1,0} - I_{2,2,2,2,0} ) \xi_0 X_{\mu } X_{\mu } X_{\nu} X_{\nu }
.
\label{eq:3.31}
\end{eqnarray}

\begin{eqnarray}
B^{(T)}_0 &=& 0
,
\nonumber \\ 
B^{(T)}_{1/2} &=& 0
,
\nonumber \\ 
B^{(T)}_1 &=&
\frac{1}{4}I_{1,2,1}\bar{\xi}_2 X_0 X_0
,
\nonumber \\ 
B^{(T)}_{3/2} &=&
\left(-\frac{1}{6} I_{1,2,0} -\frac{1}{6} I_{2,1,0}\right) 
{\bar\xi }_1 X_{\mu}  F_{0 \mu }
\nonumber \\ &&
+\left(\frac{1}{6} I_{1,2,2}-\frac{1}{6} I_{1,3,1}\right)
   \left({\bar\xi }_1 X_{0 \mu } X_{\mu }
+{\bar\xi }_1 X_{\mu 0} X_{\mu}+{\bar\xi }_1 X_{\mu \mu } X_0
-\frac{1}{2} {\bar\xi}_3 X_{0 0} X_0\right)
\nonumber \\ &&
+\left(\frac{1}{3} I_{1,1,2,2}-\frac{1}{3} I_{1,1,3,1}\right)
   \left({\bar\xi }_1 X_0 X_{\mu } X_{\mu }+{\bar\xi }_1 X_{\mu } X_0 X_{\mu}
+{\bar\xi }_1 X_{\mu } X_{\mu } X_0-\frac{1}{2} {\bar\xi}_3 X_0 X_0
   X_0\right)
,
%
\nonumber \\
B^{(T)}_2 &=&
-\frac{1}{6} I_{3,0,0} {\bar\xi}_2 F_{0\mu } F_{0\mu }
\nonumber \\ &&
+\left(\frac{1}{36}
   I_{3,2,0}-\frac{1}{2} I_{3,3,0}-\frac{1}{2} I_{4,2,0}\right)
   {\bar\xi}_2 X_{00} X_{\mu \mu }
\nonumber \\ &&
+\left(\frac{11}{36} I_{1,3,0}-\frac{1}{3}
   I_{2,2,0}-\frac{17}{36} I_{3,1,0}\right) {\bar\xi}_2 X_{0\mu } F_{0\mu
   }
\nonumber \\ &&
+\left(\frac{7}{9} I_{3,2,0}-\frac{1}{2} I_{3,3,0}-\frac{1}{2} I_{4,2,0}\right)
   {\bar\xi}_2 X_{0\mu } X_{0\mu }
\nonumber \\ &&
+\left(\frac{7}{36} I_{3,2,0}-\frac{1}{2}
   I_{3,3,0}-\frac{1}{2} I_{4,2,0}\right) {\bar\xi}_2 X_{\mu \mu
   } X_{00}
\nonumber \\ &&
+\left(-\frac{1}{2} I_{3,3,0}-\frac{1}{2} I_{4,2,0}\right)
   \left({\bar\xi}_2 X_{0\mu } X_{\mu 0}
+{\bar\xi}_2 X_{\mu 0} X_{0\mu}+{\bar\xi}_2 X_{\mu 0} X_{\mu
  ,0}-\frac{1}{2} {\bar\xi}_4 X_{00} X_{00}\right)
\nonumber \\ &&
+\left(\frac{7}{18} I_{2,1,3,0}-I_{2,1,4,0}+\frac{1}{3}
   I_{2,3,2,0}+\frac{1}{18} I_{3,1,2,0}-2 I_{3,1,3,0}-I_{4,1,2,0}\right)
   {\bar\xi}_2 X_0 X_0 X_{\mu \mu }
\nonumber \\ &&
+\left(\frac{11}{36} I_{1,1,3,0}-\frac{1}{18}
   I_{1,2,2,0}-\frac{13}{36} I_{1,3,1,0}-\frac{1}{3} I_{2,1,2,0}-\frac{11}{36}
   I_{2,2,1,0}-\frac{17}{36} I_{3,1,1,0}\right) {\bar\xi}_2 X_0 X_{\mu } F_{0\mu
   }
\nonumber \\ &&
+\left(\frac{11}{36} I_{1,1,3,0}+\frac{5}{36} I_{1,2,2,0}-\frac{11}{36}
   I_{1,3,1,0}-\frac{1}{3} I_{2,1,2,0}-\frac{1}{9} I_{2,2,1,0}-\frac{17}{36}
   I_{3,1,1,0}\right) {\bar\xi}_2 X_{\mu } X_0 F_{0\mu }
\nonumber \\ &&
+\left(\frac{7}{9}
   I_{2,1,3,0}-I_{2,1,4,0}+\frac{1}{3} I_{2,3,2,0}+\frac{7}{9} I_{3,1,2,0}-2
   I_{3,1,3,0}-I_{4,1,2,0}\right) \left({\bar\xi}_2 X_0 X_{\mu } X_{0\mu}
   +{\bar\xi}_2 X_{\mu } X_0 X_{0\mu }\right)
\nonumber \\ &&
+\left(\frac{1}{18}
   I_{2,1,3,0}-I_{2,1,4,0}+\frac{1}{3} I_{2,3,2,0}+\frac{7}{18} I_{3,1,2,0}-2
   I_{3,1,3,0}-I_{4,1,2,0}\right) {\bar\xi}_2 X_{\mu } X_{\mu
   } X_{00}
\nonumber \\ &&
+\left(-I_{2,1,4,0}+\frac{1}{3} I_{2,3,2,0}-2 I_{3,1,3,0}-I_{4,1,2,0}\right)
   \left({\bar\xi}_2 X_0 X_{\mu } X_{\mu 0}+{\bar\xi}_2 X_{\mu
   } X_0 X_{\mu 0}-\frac{1}{2} {\bar\xi}_4 X_0 X_0 X_{00}\right)
\nonumber \\ &&
+\left(\frac{5}{12} I_{2,2,2,2,0}+\frac{1}{9} I_{3,1,2,1,0}-2
   I_{3,1,3,1,0}+\frac{2}{3} I_{3,2,1,2,0}-2 I_{4,1,2,1,0}\right)
   {\bar\xi}_2 X_0 X_0 X_{\mu } X_{\mu }
\nonumber \\ &&
+\left(\frac{5}{12} I_{2,2,2,2,0}+\frac{7}{9}
   I_{3,1,2,1,0}-2 I_{3,1,3,1,0}+\frac{2}{3} I_{3,2,1,2,0}-2 I_{4,1,2,1,0}\right)
   \left(
{\bar\xi}_2 X_0 X_{\mu } X_0 X_{\mu }
+{\bar\xi}_2 X_0 X_{\mu} X_{\mu } X_0
\right.
\nonumber \\ && \hspace{1cm}
\left.
+{\bar\xi}_2 X_{\mu } X_0 X_0 X_{\mu }+{\bar\xi}_2 X_{\mu
   } X_0 X_{\mu } X_0+{\bar\xi}_2 X_{\mu } X_{\mu
   } X_0 X_0\right)
\nonumber \\ &&
+\left(-\frac{5}{24} I_{2,2,2,2,0}+I_{3,1,3,1,0}-\frac{1}{3}
   I_{3,2,1,2,0}+I_{4,1,2,1,0}\right) {\bar\xi}_4 X_0 X_0 X_0 X_0
.
\label{eq:3.33}
\end{eqnarray}

Once again we note that the $\xi_0$ and $\bar{\xi}_n$ are actually at the
left of the $I_{r_1,\ldots,r_n}$.

Further rearrangement of the expressions is possible to bring them to a more
systematic form. For instance reordering of covariant derivatives is possible
using the Bianchi identity $Y_{\mu\nu}=Y_{\nu\mu}+[F_{\mu\nu},Y]$, as well as
cyclic permutations or integration by parts. However, such extra work does not
seem to yield a simpler expression. These expressions for $B_n$ have not been
obtained directly from $A_n$ but from $\Tr\log(z-K)$ in Chan's form, to be
introduced below.

\section{Chan's form of the effective action}
\label{sec:4}

Up to now we have considered Euclidean space-times with the topologies
$\R^{\dimst}$ or $\R^{\dims}\times S^1$ appropriate to study field theories at zero or
finite temperature. The latter case leads to the Matsubara frequencies and to
the weight function $h_M(p_0)$ introduced in \eq{2.33}. At zero temperature
the weight function is just equal to unity.

As it turns out, the formalism can be carried out equally well without
assuming any particular properties of the weight function $h(p)$ in the
momentum integration. $h(p)$ can even depend on all components $p_\mu$.  For
the purpose of deriving general expressions no simplification is obtained by
imposing constraints on $h(p)$, therefore, from now on we will assume a
completely general weight function $h(p)$. We call $h$-space the setting
leading to such a weight $h(p)$ in the momentum integrals.  In the next
subsection we show that this approach does not lead to inconsistencies.

\subsection{$h$-spaces}

We devote this subsection to study the consistency of the approach with
generic $h(p)$, specifically regarding gauge invariance and cyclic property.

Generalizing the method of symbols, we define
\begin{eqnarray}
\langle x|f(D,X)|x\rangle_h &=& \int\frac{d^{\dimst}p}{(2\pi)^{\dimst}}h(p)f(D+ip,X)
,
\label{eq:4.1}
\\
\Tr_h f(D,X)  &=& \int d^{\dimst}x\,\tr\langle x|f(D,X)|x\rangle_h
= \int \frac{d^{\dimst}x\,d^{\dimst}p}{(2\pi)^{\dimst}}h(p)\,\tr f(D+ip,X)
.
\end{eqnarray}

$h(p)$ is a c-number function, therefore the cyclic property works as always:
$\Tr_h(\hat{f}_1 \hat{f}_2) = \Tr_h(\hat{f}_2 \hat{f}_1)$.\footnote{When
  $h(p)=1$, a good convergence of $f_{1,2}(D+ip,X)$ for large $p_\mu$ is
  assumed. Here we assume that this convergence is not spoiled by $h(p)$.}  As
a consequence the following property holds
\begin{equation}
\frac{\delta \Tr_h (e^K)}{\delta X(x)} = \langle x|e^K|x\rangle_h
.
\end{equation}

To extend the method of covariant symbols for generic $h(p)$ we define
\begin{equation}
h(p+iD)=
\sum_{n=0}^\infty\frac{i^n}{n!}(\partial^p_{\mu_1}\partial^p_{\mu_2}\cdots
 \partial^p_{\mu_n}h(p))D_{\mu_1}D_{\mu_2}\cdots
 D_{\mu_n}
.
\end{equation}
Then
\begin{eqnarray}
\langle x|f(D,X)|x\rangle_h
&=&
\int\frac{d^{\dimst}p}{(2\pi)^{\dimst}}h(p)e^{-iD\partial^p}
e^{iD\partial^p}f(D+ip,X)e^{-iD\partial^p}
\nonumber\\
&=&
\int\frac{d^{\dimst}p}{(2\pi)^{\dimst}} h(p+iD)f(\bar{D},\bar{X})
.
\label{eq:4.5}
\end{eqnarray}

Let us consider now the issue of gauge invariance of $\langle
x|f(D,X)|x\rangle_h$. To study this, it is convenient to write the r.h.s of 
\eq{4.1} more explicitly as
\begin{eqnarray}
\langle x|f(D,X)|x\rangle_h &=& \int\frac{d^{\dimst}p}{(2\pi)^{\dimst}}h(p)
\langle x|f(D+ip,X)|0)
.
\end{eqnarray}

Now, any operator ${\mathcal O}$ constructed with $D_\mu$ and $X(x)$
necessarily transforms gauge covariantly, i.e., as $U^{-1}{\mathcal O}
U$. Gauge covariance can be lost by taking matrix elements with the state
$|0)$, which is not covariant: in general $\langle x|{\mathcal O}|0)$ does not
transforms into $U^{-1}(x)\langle x|{\mathcal O}|0)U(x)$. However, the correct
transformation is guaranteed provided ${\mathcal O}$ is a multiplicative
operator because in this case $\langle x|{\mathcal O}|0)={\mathcal
  O}(x)\langle x|0)= {\mathcal O}(x)\to U^{-1}(x){\mathcal
  O}(x)U(x)$. Therefore, gauge covariance of $\langle x|f(D,X)|x\rangle_h$ is
ensured provided the operator
\begin{equation}
\hat{f}^\prime = \int\frac{d^{\dimst}p}{(2\pi)^{\dimst}}h(p)f(D+ip,X)
\end{equation}
is multiplicative (matrix elements $\langle x|\,|0)$ have not be taken here).
The same requirement holds for the operator $h(p+iD)$ in \eq{4.5}, namely, it
must be multiplicative. (The covariant symbol $f(\bar{D},\bar{X})$ is already
gauge covariant and multiplicative.)

For an operator ${\mathcal O}$ to be multiplicative amounts to commute with
c-number functions of $x$. This requirement can be recast in the form (the
$k_\mu$ are constant c-numbers)
\begin{equation}
e^{-ikx} {\mathcal O} e^{ikx} =  {\mathcal O}
.
\end{equation}
Due to the property $e^{-ikx} D_\mu e^{ikx} = D_\mu + ik_\mu$, we can see that
$\hat{f}^\prime$ or $h(p+iD)$ will commute with $e^{ikx}$ if 
\begin{equation}
h(p-k)=h(p)
.
\label{eq:4.9}
\end{equation}

If this condition is imposed for all $k$, the function $h(p)$ must be a
constant. This corresponds to the zero temperature case. In this case, the
quantum fields belong to the vector space $V_{\dimst}$ of arbitrary functions
of $x$ in $\R^{\dimst}$ (we disregard internal degrees of freedom here). At
finite temperature, the quantum fields are required to be periodic or
antiperiodic, and the external fields periodic. This implies that one is
working now in a subspace $V$ of $V_{\dimst}$ (namely, that of periodic or
antiperiodic functions). The operators (external fields) acting on that space
can carry only momenta of the type $k=(\vk,\omega_n)$ in order to leave $V$
invariant. Therefore, one needs to consider only this set of momenta when
checking the relation $h(p-k)=h(p)$ for $h(p)=h_M(p_0)$ (and the relation is
of course fulfilled by $h_M(p_0)$.) At the same time, the restriction in $k$
is directly related with the compactification $\R^{\dimst}\to\R^{\dims}\times
S^1$.

Let us generalize these ideas for other $h(p)$.  There should be a vector
space $V$ of space-time functions for the quantum fields, a set $A$ of allowed
operators leaving $V$ invariant, and a set $K$ of allowed momenta. The
operators in $A$ are those having only momenta $k$ in $K$ in their
decomposition in Fourier modes. Because combinations of operators in $A$
should also stay in $A$, we must demand that if $k_1,k_2\in K$, $k_1\pm k_2\in
K$ (i.e., the set $K$ is closed under linear combination with integer
coefficients, in particular $0\in K$). On the other hand, $V$ is composed of
those functions with Fourier modes of the type $q+k$, for some fixed $q$ and
$k\in K$. Ideally, such $K$ would come from some suitable compactification of
$\R^{\dimst}$. Finally, there will be gauge invariance provided $h(p-k)=h(p)$
for all $p$ and all $k$ in $K$.

In practice the only obvious setting carrying out the above program is for
space-times of the type $\R^n\times S^1\times \cdots\times S^1$, $0\le n\le
\dimst$. This corresponds to modes $k$ which are an integer linear combination
of $\dimst-n$ fixed linearly independent vectors plus an arbitrary vector in
the $n$ supplementary directions. In this case $h(p+iD)$ is a function of the
$\dimst-n$ ``Polyakov loops'' in the $\dimst-n$ compactified directions.

At present, it is not clear whether there exist other useful realizations of
$h$-spaces. In any case, the formalism can be developed without special
assumptions on $h(p)$. In what follows we simply assume that the quantum
fields lie in the appropriate space $V$ (the $h$-space) and the allowed
external fields, as well as the allowed gauge transformations, leave $V$
invariant.

\subsection{$X$-form and $N$-form of the expressions}

\subsubsection{Diagonal matrix elements of the propagator}

Let the propagator be
\begin{equation}
G(z) = \frac{1}{z-K}
.
\end{equation}
As is well-known one can obtain generic functions of $K$ from the propagator,
\begin{equation}
f(K)= \int_\Gamma \frac{dz}{2\pi i} f(z) G(z)
,
\end{equation}
where $\Gamma$ encloses counterclockwise the spectrum of $K$. ($f(z)$ is
assumed to have the required good properties.)

The diagonal matrix elements of the propagator in the $h$-space can be
computed using the method of symbols or covariant symbols and the derivative
expansion, as already explained for the heat kernel. To second order one finds:
\begin{eqnarray}
\langle x|G(z)|x\rangle_h &=& 
\int\frac{d^{\dimst}p}{(2\pi)^{\dimst}}h(p+iD)\Big(
N
-2 i p_{\mu } N X_{\mu } N^2
-4 p_{\mu } p_{\nu } N X_{\mu  \nu } N^3 + N X_{\mu  \mu } N^2
\nonumber\\ &&
-8 p_{\mu } p_{\nu } N X_{\mu } N X_{\nu } N^3
-4 p_{\mu } p_{\nu } N X_{\mu} N^2 X_{\nu } N^2
+2 N X_{\mu } N X_{\mu } N^2
+O(D^3)
\Big)
.
\label{eq:4.12}
\end{eqnarray}
Here
\begin{equation}
N= (z+p^2-X)^{-1}
.
\end{equation}

The expression through third order is given in Appendix \ref{app:C}, using
labeled operators.

We refer to the form in \eq{4.12} as the {\em $X$-form of the expression} because
the $X$ appear with derivatives and the $N$ carry no derivative.  By means of
the relation $X_\mu=N^{-1}N_\mu N^{-1}$, and derivatives of it, one can
eliminate completely the $X$ and write the same expression using only $N$ and
covariant derivatives of it.  For a generic initial expression, negative
powers of $N$ will be present after elimination of $X$. When this is not the
case we say that the expression {\em admits an $N$-form}. As it turns out, the
covariant symbol of the propagator admits an $N$-form (see Appendix
\ref{app:C}).\footnote{We do not have a proof of this to all orders (but have
  little doubt that it is so). It has been verified through fourth order in
  the derivative expansion.} As a consequence, the diagonal matrix element of
the propagator also admits an $N$-form. One virtue of the $N$-form is that
usually the expressions are much more compact. A drawback is that the
functions $I_{r_1,\ldots,r_n}$ do not directly apply for expressions written
in $N$-form.

For the diagonal matrix elements of the propagator, through third order in the
derivative expansion and  in $N$-form, one finds:
\begin{eqnarray}
\langle x|G(z)|x\rangle_h
&=&
\int\frac{d^{\dimst}p}{(2\pi)^{\dimst}}h(p+iD)\Big(
N
-2 i p_{\mu } N_{\mu   } N
-4 p_{\mu } p_{\nu } N_{\mu } N_{\nu } N
-4 p_{\mu } p_{\nu } N_{\mu \nu } N^2
+N_{\mu  \mu } N
\nonumber \\ &&
-2 i   p_{\mu } N_{\mu  \nu } N_{\nu } N
-2 i p_{\mu } N_{\nu  \mu } N_{\nu } N
-2 i p_{\mu } N_{\mu } N_{\nu  \nu } N
-2 i p_{\mu   } N_{\nu  \nu } N_{\mu } N
\nonumber \\ &&
-\frac{4}{3} i p_{\mu } N_{\mu  \nu  \nu } N^2
-\frac{4}{3} i   p_{\mu } N_{\nu  \mu  \nu } N^2
-\frac{4}{3} i p_{\mu } N_{\nu \nu \mu } N^2
-2i p_{\mu } N F_{\mu  \nu } N_{\nu } N
-\frac{2}{3} i   p_{\mu } N F_{\nu  \mu  \nu } N^2
\nonumber \\ &&
+8 i p_{\mu } p_{\nu } p_{\alpha } N_{\mu } N_{\nu    \alpha } N^2
+8 i p_{\mu } p_{\nu } p_{\alpha } N_{\mu  \nu } N N_{\alpha } N
+16 i   p_{\mu } p_{\nu } p_{\alpha } N_{\mu  \nu } N_{\alpha } N^2
+8 i p_{\mu } p_{\nu   } p_{\alpha } N_{\mu  \nu  \alpha } N^3
\nonumber \\ &&
+8 i p_{\mu } p_{\nu } p_{\alpha } N_{\mu} N_{\nu } N_{\alpha } N
%
+O(D^4)
\Big)
.
\label{eq:4.12b}
\end{eqnarray}
The corresponding expression for the fourth order terms is given in Appendix
\ref{app:C}. In these expressions there are no ambiguities related to
integration by parts in $p_\mu$ or $z$ and so the formulas are essentially
unique. The only remaining freedom is to reorder the covariant derivatives.

\subsubsection{Trace of the propagator}

In order to obtain the trace of a generic function of $K$, one can use
\begin{equation}
\Tr_h\, f(K)= \int_\Gamma \frac{dz}{2\pi i} f(z) \,\Tr_h \,G(z)
.
\label{eq:4.15}
\end{equation}

The expression of $\Tr_h \,G(z)$ can be obtained by starting from $\langle
x|G(z)|x\rangle_h$ (\eq{4.12b}) and using integration by parts and the trace
cyclic property to obtain a simpler form.  
Due to the presence of the factor $h(p+iD)$, the integration by parts (with
respect to the covariant derivative) and the cyclic property do not act in the
usual way for the expression in parenthesis. Instead one can use the identity
\begin{equation}
\int\frac{d^{\dimst}p}{(2\pi)^{\dimst}}
A(p)h(p+iD) B(p)
=
\int\frac{d^{\dimst}p}{(2\pi)^{\dimst}}
h(p+iD) e^{i\partial^p\hat{D}_A} A(p) B(p)
.
\label{eq:4.16}
\end{equation}
Here $A(p)$ and $B(p)$ are arbitrary operators which may depend on $p_\mu$
(but not on $\partial^p_\mu$). $\hat{D}_{A,\mu}$ is $[D_\mu,~]$ acting only on
$A(p)$. On the other hand, $\partial^p_\mu$ acts on the $p_\mu$ dependence in
$A(p)$ and $B(p)$. This identity is proven in Appendix \ref{app:D}. Of course,
if one is working modulo $O(D^{n+1})$, and $AB=O(D^n)$, the operator
$e^{i\partial^p\hat{D}_A}$ can be dropped, and the cyclic property works as
usual.

However, the expression for $\Tr_h G(z)$ is more easily obtained from
the relation
\begin{equation}
\Tr_h G(z) = \frac{d}{dz} \Tr_h\log(z-K)
,
\end{equation}
using the compact expression for $\Tr_h\log(z-K)$ to be given below
(\eq{4.23}).  An explicit calculation to third order gives:
\begin{eqnarray}
\Tr_h \,G(z)
&=&
\int\frac{d^{\dimst}x\,d^{\dimst}p}{(2\pi)^{\dimst}}\,\tr\!\Big[h(p+iD)\Big(
N
-4p_{\mu}p_{\nu}NN_{\mu}N_{\nu}
-6ip_{\mu}N_{\mu}N_{\nu}N_{\nu}
\nonumber \\ &&
+ip_{\mu}F_{\mu\nu}NNN_{\nu}
+ip_{\mu}F_{\mu\nu}NN_{\nu}N
+\frac{2}{3}ip_{\mu}F_{\mu\nu}N_{\nu}NN
\nonumber \\ &&
+\frac{8}{3}ip_{\mu}p_{\nu}p_{\alpha}NNN_{\mu}N_{\nu\alpha}
+10ip_{\mu}p_{\nu}p_{\alpha}NNN_{\mu\nu}N_{\alpha}
+\frac{26}{3}ip_{\mu}p_{\nu}p_{\alpha}NN_{\mu}NN_{\nu\alpha}
\nonumber \\ &&
-2ip_{\mu}NN_{\nu\mu}N_{\nu}
-2ip_{\mu}NN_{\nu\nu}N_{\mu}
+44ip_{\mu}p_{\nu}p_{\alpha}NN_{\mu}N_{\nu}N_{\alpha}
+O(D^4)
\Big)
\Big]
.
\label{eq:4.18}
\end{eqnarray}

Also the matrix elements of the propagator can be recovered from the logarithm
by using
\begin{equation}
\langle x|G(z)|x\rangle_h
=
-\frac{\delta}{\delta X(x)}
\Tr_h\log(z-K)
,
\end{equation}
but in this case the relation \eq{4.16} is needed to extract the factor
$\delta X(x)$ in $\delta\Tr_h\log(z-K)$.

\subsection{The effective action in Chan's form}

As follows from \eqs{4.15} and (\ref{eq:4.18}), for a generic function of $K$,
$\Tr_h f(K)$ requires an integral over $p_\mu$ and another over $z$.
Nevertheless, the parametric integration over $z$ can be obviated in the
special case of the logarithm. $\Tr_h\log K$ is just the effective action.

\subsubsection{$\Tr_h \log(z- K)$ }

As it turns out (verified through four derivatives) the diagonal matrix
elements of the propagator can be written as
\begin{equation}
\langle x|G(z)|x\rangle_h
=
\int\frac{d^{\dimst}p}{(2\pi)^{\dimst}}\Big[
 h(p+iD) \frac{d{\mathcal M}(z) }{dz}
+ 
h(p) {\mathcal C}(z) 
\Big]
.
\end{equation}
Here 
${\mathcal M}(z)$ is a multiplicative operator that admits an $N$-form, and
${\mathcal C}(z)$ is traceless (a sum of commutators). Therefore,
\begin{eqnarray}
\Tr_h\log(z-K) &=&
\int_\Gamma \frac{d\zeta}{2\pi i} \log(z-\zeta)
\int\frac{d^{\dimst}x\,d^{\dimst}p}{(2\pi)^{\dimst}}\,\tr\!\Big[h(p+iD)
\frac{d{\mathcal M}(\zeta)}{d\zeta}
\Big].
\end{eqnarray}
The term with ${\mathcal C}(z)$ has dropped from the expression.  Next we
integrate by parts in $\zeta$, this transforms $\log(z-\zeta)$ into
$1/(z-\zeta)$.  The integrand is assumed to be well behaved at infinity (in
particular, the branch cut of the logarithm is no longer present). Hence, we
can switch from the contour $\Gamma$, that includes the spectrum of $K$ and
excludes the pole at $\zeta=z$, to a contour excluding the spectrum of $K$ and
including the pole at $\zeta=z$. This produces
\begin{eqnarray}
\Tr_h\log(z-K) &=&
\int\frac{d^{\dimst}x\,d^{\dimst}p}{(2\pi)^{\dimst}}\,\tr\!\Big[h(p+iD) 
{\mathcal M}(z)
  \Big].
\label{eq:4.22}
\end{eqnarray}

Now $\Tr_h\log(z-K)$ is written in {\em Chan's form}, namely, in $N$-form and
without parametric integration on $\zeta$. Note that the dependence on $z$ is
inessential as $z$ can be absorbed in $X$.

Explicitly,
\begin{eqnarray}
{\mathcal M}(z)
&=&
-\log N+p_{\mu}p_{\nu}N_{\mu}N_{\nu}
\nonumber\\&&
-\frac{1}{3}ip_{\mu}NN_{\nu}F_{\mu\nu}
-\frac{1}{3}ip_{\mu}N_{\nu}NF_{\mu\nu}
-\frac{2}{3}ip_{\mu}p_{\nu}p_{\alpha}N_{\mu\nu}NN_{\alpha}
+\frac{2}{3}ip_{\mu}p_{\nu}p_{\alpha}N_{\mu\nu}N_{\alpha}N
\nonumber\\&&
-\frac{1}{4}N_{\mu\mu}N_{\nu\nu}
+\frac{1}{2}N_{\mu}N_{\nu}F_{\mu\nu}
+\frac{1}{12} N^2 F_{\mu\nu}F_{\mu\nu}
\nonumber\\&&
+\frac{1}{9}p_{\mu}p_{\nu}N_{\mu\nu}N_{\alpha\alpha}N
+\frac{7}{9}p_{\mu}p_{\nu}N_{\alpha\alpha}N_{\mu\nu}N
+\frac{28}{9}p_{\mu}p_{\nu}N_{\mu\alpha}N_{\nu\alpha}N
\nonumber\\&&
-\frac{17}{9}p_{\mu}p_{\nu} N^2 N_{\mu\alpha}F_{\nu\alpha}
-\frac{4}{3}p_{\mu}p_{\nu}NN_{\mu\alpha}NF_{\nu\alpha}
+\frac{11}{9}p_{\mu}p_{\nu}N_{\mu\alpha} N^2 F_{\nu\alpha}
\nonumber\\&&
-\frac{11}{9}p_{\mu}p_{\nu}N_{\alpha}NN_{\mu}F_{\nu\alpha}
-\frac{11}{9}p_{\mu}p_{\nu}NN_{\mu}N_{\alpha}F_{\nu\alpha}
-\frac{4}{9}p_{\mu}p_{\nu}NN_{\alpha}N_{\mu}F_{\nu\alpha}
\nonumber\\&&
-\frac{13}{9}p_{\mu}p_{\nu}N_{\mu}NN_{\alpha}F_{\nu\alpha}
-\frac{2}{9}p_{\mu}p_{\nu}N_{\mu}N_{\alpha}NF_{\nu\alpha}
+\frac{5}{9}p_{\mu}p_{\nu}N_{\alpha}N_{\mu}NF_{\nu\alpha}
\nonumber\\&&
-\frac{2}{3}p_{\mu}p_{\nu} N^3 F_{\mu\alpha}F_{\nu\alpha}
\nonumber\\&&
+\frac{8}{3}p_{\mu}p_{\nu}p_{\alpha}p_{\beta}N_{\mu}N_{\nu\alpha}N_{\beta}N
-4p_{\mu}p_{\nu}p_{\alpha}p_{\beta}N_{\mu\nu}NN_{\alpha\beta}N
-4p_{\mu}p_{\nu}p_{\alpha}p_{\beta}N_{\mu\nu}N_{\alpha\beta} N^2
\nonumber\\&&
+\frac{10}{3}p_{\mu}p_{\nu}p_{\alpha}p_{\beta}N_{\mu}N_{\nu}N_{\alpha}N_{\beta}
+O(D^5)
.
\label{eq:4.23}
\end{eqnarray}
(The isolated term $-\log N= \log (N^{-1})$ is still considered to be in
$N$-form.)

The form of ${\mathcal M}(z)$ is not unique, due to the cyclic property and
integration by parts with respect to the covariant derivative.

That Chan's form exists is not trivial, in the sense that it holds for the
logarithm but not for generic functions of $K$. Chan's form was introduced in
\cite{Chan:1986jq}. Extended to six derivatives in \cite{Caro:1993fs}, to
curved space-time in \cite{Salcedo:2007bt}, and to fermions in
\cite{Salcedo:2008bs}. It is quite remarkable that it also exists in
$h$-spaces (in particular, at finite temperature). This is more so as we are
not allowed to use two important tools of the original derivation by Chan
\cite{Chan:1986jq}, namely, momentum average and integration by parts with
respect to $p_\mu$. This is forbidden due to the presence of the function
$h(p)$, which is arbitrary. It is noteworthy that, unlike the original Chan's
formula, our expression does not depend on the space-time dimension. This
property is also shared by the heat kernel. Another difference with Chan's
result is that the $p_\mu$ are contracted only with covariant derivative
indices and not with other $p_\mu$.

\subsubsection{Traced heat kernel}

To obtain the traced heat kernel, \eq{3.27}, from the effective action,
eqs. (\ref{eq:4.22}) and (\ref{eq:4.23}), one can use
\begin{eqnarray}
\Tr_h e^K &=& 
\int_\Gamma \frac{dz}{2\pi i}e^z \Tr_h\frac{1}{z-K} 
\nonumber \\
&=&
\int_\Gamma \frac{dz}{2\pi i}e^z \frac{\partial}{\partial z}
\Tr_h\log(z-K) 
\nonumber \\
&=&
-\int_\Gamma \frac{dz}{2\pi i}e^z \Tr_h\log(z-K)
\nonumber \\
&=&
-\int_\Gamma \frac{dz}{2\pi i}e^z
\int\frac{d^{\dimst}x\,d^{\dimst}p}{(2\pi)^{\dimst}}\,\tr\!\Big[h(p+iD) {\mathcal M}(z)
  \Big]
.
\end{eqnarray}
Now the shift $z\to z - p^2$ implies $N\to (z-X)^{-1}$ in ${\mathcal M}(z)$,
and $e^z\to e^z e^{-p^2}$. Hence, ${\mathcal M}(z)$ becomes $p$-independent
and the integral over momenta reduces to obtaining the following $h$-dependent
operators
\begin{equation}
\langle p_{\mu_1}\cdots p_{\mu_n}\rangle_h
=
(4\pi)^{\dimst/2}
\int\frac{d^{\dimst}p}{(2\pi)^{\dimst}} h(p+iD) e^{-p^2} p_{\mu_1}\cdots p_{\mu_n}
.
\end{equation}
The $B_n$ in section \ref{subsec:IIIB} are obtained in this way.

\subsubsection{Reduction to Chan's form}
In what follows, we explain how \eq{4.23} is obtained. First, let us see how
Chan's derivation \cite{Chan:1986jq} can be adapted to the present case. Using
\eq{4.1},
\begin{eqnarray}
\langle x| \log(z-K)|x\rangle_h
&=&
\int \frac{d^{\dimst}p}{(2\pi)^{\dimst}}\,h(p)\log\big(z-(D_\mu+ip_\mu)^2-X\big)
\nonumber \\
&=&
\int \frac{d^{\dimst}p}{(2\pi)^{\dimst}}\,h(p)
\left[
\log\big(N^{-1}) + \log(1-(2ip_\mu D_\mu + D_\mu^2)N\big)
+{\mathcal C}
\right]
\nonumber \\
&=&
\int \frac{d^{\dimst}p}{(2\pi)^{\dimst}}\,h(p)
\left[
\log(N^{-1})-
\sum_{n=1}^\infty\frac{1}{n}
\Big((2ip_\mu D_\mu + D_\mu^2)N\Big)^n
+{\mathcal C}
\right]
,
\label{eq:4.24}
\end{eqnarray}
where ${\mathcal C}$ denote commutator terms, which will vanish upon use of the
cyclic property of trace. To second order in the derivative expansion
\begin{eqnarray}
\Tr_h \log(z-K)
&=&
\int \frac{d^{\dimst}x\,d^{\dimst}p}{(2\pi)^{\dimst}}\tr\Big[
h(p)\Big(
\log(N^{-1})
-2ip_\mu D_\mu N -D_\mu^2 N
+2p_\mu p_\nu D_\mu N  D_\nu N
+O(D^3) 
\Big)
\Big]
.
\end{eqnarray}
Using the relations
\begin{eqnarray}
\partial^p_\mu\log (N^{-1}) &=& 2p_\mu N
,
\nonumber\\ 
\frac{1}{2}\partial^p_\mu \partial^p_\nu \log (N^{-1}) 
&=&
\delta_{\mu\nu}N - 2p_\mu p_\nu N^2
,
\end{eqnarray}
the trace can be written as
\begin{eqnarray}
\Tr_h \log(z-K)
&=&
\int \frac{d^{\dimst}x\,d^{\dimst}p}{(2\pi)^{\dimst}}\tr\Big[
h(p)\Big(
\log(N^{-1})
-iD_\mu\partial^p_\mu \log(N^{-1})
-\frac{1}{2}D_\mu D_\nu\partial^p_\mu\partial^p_\nu \log(N^{-1})
\nonumber \\ &&
-2 p_\mu p_\nu D_\mu D_\nu N^2
+2 p_\mu p_\nu D_\mu N D_\nu N
+O(D^3) 
\Big)
\Big]
\nonumber \\
&=&
\int \frac{d^{\dimst}x\,d^{\dimst}p}{(2\pi)^{\dimst}}\tr\Big[
h(p)e^{-iD\partial^p}
\Big(
\log(N^{-1})
+p_\mu p_\nu N_\mu N_\nu
+O(D^3) 
\Big)
\Big]
\nonumber \\
&=&
\int \frac{d^{\dimst}x\,d^{\dimst}p}{(2\pi)^{\dimst}}\tr\Big[
h(p+iD)
\Big(
\log(N^{-1})
+p_\mu p_\nu N_\mu N_\nu
+O(D^3) 
\Big)
\Big]
.
\end{eqnarray}
This expression has the desired Chan's form.

In order to obtain the expression of $\mathcal M(z)$ to four derivatives it is
not practical to apply the previous method since it is not sufficiently
systematic. A possibility would be to simply write down all possible terms
that could appear in $\mathcal M(z)$ to fourth order, with free coefficients,
and expand everything in powers of $D_\mu$, including $h(p+iD)\to
h(p)e^{-iD\partial^p}$, using the cyclic property, to match the terms in
\eq{4.24}. Assuming that the $p_\mu$ can be only contracted with covariant
derivatives (but not with other $p_\mu$) the number of terms is finite (since
$N^{-1}$ is not allowed). However, the number of possible terms is too large
(and it is easy to miss some of them when trying to write down all of terms).

The method that we have followed is partially constructive and partially
guessing. Let 
\begin{equation}
{\mathcal A}(z)
=
\log(N^{-1})-
\sum_{n=1}^\infty\frac{1}{n}
\Big((2ip_\mu D_\mu + D_\mu^2)N\Big)^n
+{\mathcal C}
,
\label{eq:4.28}
\end{equation}
where ${\mathcal C}$ are suitable commutator terms to be fixed. From previous
formulas,
\begin{eqnarray}
\Tr_h \log(z-K)
&=&
\int \frac{d^{\dimst}x\,d^{\dimst}p}{(2\pi)^{\dimst}}\tr\!\!\left[
h(p)
{\mathcal A}(z)
\right]
\nonumber\\ 
&=&
\int \frac{d^{\dimst}x\,d^{\dimst}p}{(2\pi)^{\dimst}}\tr\!\!\left[
h(p+iD)
e^{iD\partial^p}
{\mathcal A}(z)
e^{-iD\partial^p}
\right]
.
\end{eqnarray}
Hence, we have to choose ${\mathcal C}$, if possible, in such a way that the
operator
\begin{equation}
{\mathcal M}(z) = e^{iD\partial^p}
{\mathcal A}(z)
e^{-iD\partial^p}
\label{eq:4.30}
\end{equation}
is multiplicative and in $N$-form. To see how this condition reflects on
${\mathcal A}(z)$, let us define two first-order variations, namely,
\begin{eqnarray}
\delta_D &:& \quad D_\mu \to D_\mu +i\delta a_\mu 
,
\nonumber \\
\delta_p &:& \quad p_\mu \to p_\mu +\delta a_\mu 
,
\end{eqnarray}
where $\delta a_\mu$ is an arbitrary constant c-number (common to both
variations). Clearly, the condition that ${\mathcal M}(z)$ is multiplicative
(and so with the covariant derivative operators in the form $[D_\mu, ~]$)
is that
\begin{equation}
\delta_D {\mathcal M}(z)  = 0
.
\end{equation}
Using \eq{4.30}, this requirement translates into the following condition on
${\mathcal A}(z)$:
\begin{equation}
(\delta_D-\delta_p) {\mathcal A}(z)  = 0
.
\label{eq:4.34}
\end{equation}
In turn, this is just the condition requiring that ${\mathcal A}(z)$ must
depend only on the combination $D_\mu+ip_\mu$. This property is manifest in
the symbol $\log(z-(D_\mu+ip_\mu)^2-X)$, but is not automatically preserved by
the derivative expansion with formal use of the cyclic property (which is
needed to have an $N$-form). So we have to choose the freedom implied by the
cyclic property (i.e., the commutator terms ${\mathcal C}(z)$) to fulfill
\eq{4.34}.

What we have done is to expand ${\mathcal A}(z)$ in \eq{4.28}, but allowing
all possible cyclic permutations for each term, with free coefficients (this
is the guess). Such coefficients are then partially fixed by the condition of
reproducing $\log(z-(D_\mu+ip_\mu)^2-X)$, modulo the cyclic property, and by
the condition in \eq{4.34}. This condition is easily implemented by means of
the rules
\begin{equation}
(\delta_D-\delta_p) D_\mu = i\delta a_\mu
,
\quad
(\delta_D-\delta_p) p_\mu = -\delta a_\mu
,
\quad
(\delta_D-\delta_p) N = 2\delta a_\mu p_\mu N^2
.
\end{equation}
The corresponding ${\mathcal M}(z)$ obtained from \eq{4.30} is
multiplicative. It can be written in a manifestly multiplicative form by
moving the $D_\mu$ to the right, forming covariant derivatives. The remaining
freedom in the coefficients is used to obtain a simple form for ${\mathcal
  M}(z)$. The guess chosen works at least to four derivatives, and very likely
also to all orders. We conjecture that Chan's form for general $h(p)$ can 
be extended to curved space-times as well.

\subsubsection{Quenched approximation}
\label{subsubsec:IIIC4}

The Polyakov loop in the formulas, or more generally the explicit $iD_\mu$ in
$h(p+iD)$, is needed by consistency, but it is also a nuisance. Here we study
the effect of setting this explicit $iD_\mu$ to zero by hand in an
expression. We call this the quenched version of the expression. The quenched
results will be incorrect in general, but still one can consider whether this
approximation can be done consistently.

For any operator, the same derivation leading to \eq{4.5} can be repeated
putting the explicit $iD_\mu$ to the right:
\begin{eqnarray}
\langle x|f(D,X)|x\rangle_h
&=&
\int\frac{d^{\dimst}p}{(2\pi)^{\dimst}} h(p+iD)f(\bar{D},\bar{X})
=
\int\frac{d^{\dimst}p}{(2\pi)^{\dimst}} f(\bar{D},\bar{X}) h(p+iD)
.
\end{eqnarray}
After quenching, by setting $iD_\mu$ to zero, the two expressions yield
two different (incorrect) results. In fact, for a Hermitian operator like
$e^K$, the unquenched matrix element respects hermiticity, but the two quenched
expressions do not (rather they are hermitian conjugate of each other). On the
other hand, inside the trace, the two quenched expressions do coincide (with
each other, but not with the exact one containing $h(p+iD)$)\footnote{Inside
  the trace and the integral over $p_\mu$, $ h(p)f(\bar{D},\bar{X})=
  h(p)e^{i\partial^p D}f(D+ip,X)= h(p-iD)f(D+ip,X)= f(D+ip,X) h(p-iD)
  =f(\bar{D},\bar{X})h(p) $.}
\begin{equation}
\Tr_{h,q}(f(D,X)) =
\int\frac{d^{\dimst}x d^{\dimst}p}{(2\pi)^{\dimst}}\tr\!\!\left[
  h(p)f(\bar{D},\bar{X})
\right]
=
\int\frac{d^{\dimst}x d^{\dimst}p}{(2\pi)^{\dimst}} \tr\!\!\left[
f(\bar{D},\bar{X}) h(p)\right]
.
\label{eq:4.38}
\end{equation}
This relation provides a concrete choice of quenched version of the trace of
an operator $f(D,X)$. In general this will not coincide with first computing
$\Tr_h(f(D,X))$ and then quenching, since the latter does not commute with the
cyclic property or integration by parts.

Next, we study whether the quenched version of the traced heat kernel (as
defined from \eq{4.38} using $e^K$) satisfies a consistency condition like
that in eqs. (\ref{eq:3.27}) and (\ref{eq:3.29}), namely,
\begin{equation}
\langle x|e^K|x\rangle_{h,q} =
\frac{\delta \Tr_{h,q} (e^K)}{\delta X(x)}
,\qquad
\Tr_{h,q}(e^K) = 
\int d^{\dimst}x \,\tr\langle x|e^K|x\rangle_{h,q}
\,.
\end{equation}
This kind of conditions, and similar ones for the effective action, can be
derived from the corresponding relation for the propagator (which of course
holds in the unquenched case too)
\begin{equation}
\int d^{\dimst}x \,\tr \frac{\delta \Tr_{h,q} G(z)}{\delta X(x)} =
-\frac{d}{dz}\Tr_{h,q}G(z) ,
\label{eq:4.40}
\end{equation}
As a matter of fact, \eq{4.40} is correct, just because the expressions admit
an $N$-form and so depend on $z$ and $X$ in the form $z-X$. (Note that $z-X$
appears only in $N$, and any given block $N_{\mu_1\ldots\mu_n}$ can be
transformed into $N$ by integration by parts. So one needs to consider only
the $X$ and $z$ variations on each block $N$ at a time. The two variations
give the same result thanks to the trace.)

The quenched version of the effective action, as defined in \eq{4.38} using
$\log K$, does
not admit a Chan's form. This can be seen from \eq{4.12b}. (Note that
\eq{4.18} has already applied the cyclic property and the integrand shown
there is not unique.) After applying the quenching prescription in \eq{4.12b},
one finds that the term $N_{\mu\mu} N$ cannot be expressed as a derivative
with respect to $z$, modulo commutator terms.

We stress once more that, in general, setting $iD_\mu=0$ in a traced quantity,
written in different ways related by the cyclic property, yields different
results. This follows from \eq{4.16}. At finite temperature this is also clear
from \eq{2.18}: when using the cyclic property, commutation with $\Omega$
produces time derivatives which are missed if $\Omega$ is set to unity by
hand. Also, relevant contributions can be missed by quenching. For instance,
the first contribution to the induced charge density, obtained by taking a
variation with respect to the potential $A_0(x)$ in the effective action,
comes from the Polyakov loop. Other contributions coming from $F_{\mu\nu}(x)$
contain more derivatives.

All this implies that the quenched approximation is rather dangerous and may
produce uncontrolled results. Therefore, quenching should either be avoided
altogether or at least, a careful evaluation of the contribution coming from
the terms neglected should be done.

\section{Summary and conclusions}
\label{sec:5}

We have developed a new technique to deal with diagonal matrix elements of
generic pseudo-differential operators. The method applies at finite
temperature, or more generally, to $h$-spaces, i.e., spaces with weighted
integrals over the momentum of the loop. The approach is based on extending
the method of covariant symbols to such spaces. This allows to carry out a
manifestly gauge covariant and Lorentz covariant calculation throughout. We
conjecture that the approach can be extended to curved space-time as well. In
this case, the Polyakov loop of the Levi-Civita connection is expected to
appear in parallel with the gauge connection.

The new technique is appropriate to carry out covariant derivative expansions,
so we have applied it to the heat kernel and to the effective action in Chan's
form. For the heat kernel we present results for the diagonal matrix elements
to three derivatives (the fourth order terms have also been obtained but are
too bulky to be included). For the trace of the heat kernel we present results
to four derivatives. We also present, to four derivatives, the expression of
the effective action of a generic bosonic Klein-Gordon operator in Chan's form
(i.e., prior to momentum integration) valid in $h$-spaces.

We have briefly touched the connection with the real time formulation of field
theory at finite temperature. That formulation is appropriate to treat time
dependent aspects, or even problems related to non-equilibrium physics. Such
connection, not made in this work, is clearly worth pursuing.

In this regard, we emphasize that the use of generic weights $h(p)$ in this
work is not intended as a device to describe time dependent situations or a
connection to the real time formalism.\footnote{Nevertheless, the possibility
  of transforming \eq{2.44} into \eq{2.48} suggests that the use of weights
  $h(p)$ defined on complex paths could have a bearing on this
  subject. Successful use of complex weights along complex paths in hadronic
  physics can be seen, for instance, in the spectral quark model introduced in
  \cite{RuizArriola:2003bs}. } We merely observed that, although $h_M(p_0)$ is
an even function of $p_0$, all the Chan-like formulas can be written equally
well for any $h(p)$ without assuming any parity property or, more generally,
any special dependence on $p_\mu$. We find this formal property remarkable.
However, as shown in \eq{4.9}, the condition of gauge invariance does
introduce a requirement of periodicity, which in practice we only know how to
fulfill for Euclidean space-times with the topology of a (possibly
degenerated) torus.  It could be that the formal property is just a
mathematical nicety, or it could signal a deeper property of the
formalism. This is not known at present.

As emphasized in Section \ref{sec:2} the derivative expansion implies a
resummation of the dimensional (or large mass) expansion and in this sense we
go considerably further that the standard approach for the heat kernel.  The
derivative expansion has the virtue of being gauge covariant at each order
separately. In addition, higher order terms are increasingly ultraviolet
convergent. So, for instance, anomaly saturating effective actions can be
computed in closed form using this technique
\cite{Salcedo:1996qy,Salcedo:2002pr}. The derivative expansion in field theory
is one of the few systematic tools to go beyond the perturbative regime
\cite{Ball:1989xg,Avramidi:1990je,Schubert:2001he}, and provides guidance in
modeling of effective Lagrangians in exact renormalization group approaches
\cite{Pawlowski:2005xe}. The applicability of this technique extends to
external field configurations which are slowly varying on space-time.  In the
case of external gauge fields, these have to be weak since they enter in the
covariant derivative to preserve gauge invariance. The range of validity of
the derivative expansion as compared to exact calculations in concrete
profiles have been tested, most recently, in \cite{Dunne:2011fp}. As expected
the truncated expansion works better for quantum fields with short
wavelengths, although even outside this regime it does a good job in some of
the cases studied \cite{Dunne:2011fp}. In general the derivative expansion is
expected to be asymptotic, like the semiclassical expansion in quantum
mechanics, with which it is closely related. Therefore a naive summation of
higher orders would not provide a convergent result. To carry out the
expansion to fourth order is rather standard, as the number of terms quickly
increases beyond that order \cite{Caro:1993fs,Salcedo:2004yh}. In addition, in
four space-time dimensions, the calculation to four derivatives accounts for
the ultraviolet divergent contributions. These are the ones leading to power
counting renormalizable Lagrangian terms.

Matrix elements of operators acting in the one-particle Hilbert space,
$\langle y|A|x\rangle$, correspond to propagation along single lines from $x$
to $y$ in the Feynman graphs (in position space). General Feynman diagrams can
be constructed joining these lines with the vertices of the theory under
consideration. In the present work we have only studied diagonal matrix
elements $\langle x|A|x\rangle$. This is appropriate to produce local
Lagrangians, under the derivative expansion. However, such restriction is
certainly a limitation. For instance, as noted in Section \ref{sec:2A}, we can
compute in this way the expectation value of the induced current $\langle
J_\mu(x)\rangle$. By the same token, sum rules of the type $\langle T (D^n
J(x)) (D^m J(x))\rangle$ can be also obtained from the local calculation. On
the other hand, the correlation between two currents in two different points
$\langle T J_\mu(x) J_\nu(y)\rangle$ is not directly accessible. Therefore, an
extension of the techniques discussed here to non diagonal matrix elements
would be of interest.

Upon completion of this work, we have learned that an equation equivalent to
our \eq{2.36} has been found independently in \cite{Brauner:2011vb} in
their study of CP violation in the standard model at finite temperature.

\begin{acknowledgments}
Research supported by DGI (FIS2008-01143), Junta de Andaluc{\'\i}a grant
FQM-225, the Spanish Consolider-Ingenio 2010 Programme CPAN (CSD2007-00042),
and it is part of the European Community-Research Infrastructure Integrating
Activity Study of Strongly Interacting Matter (acronym HadronPhysics2, Grant
Agreement n. 227431), under the Seventh EU Framework Programme. The authors
benefited from exchange of ideas by the ESF Research Network CASIMIR.
\end{acknowledgments}

\newpage
\clearpage

\appendix
\section{Commutator expansion}
\label{app:A}

If an expansion can be defined by means of a bookkeeping parameter the
corresponding coefficients are well defined: $f(\lambda)=\sum_n c_n\lambda^n$
and $c_n$ does not depend on how the expression is manipulated. Unfortunately,
this is not the case for expansions based on counting the number of
commutators. For instance, consider operators $\hat{f}=f(A,B)$ given by linear
combinations of products of the basic operators $A$ and $B$. No particular
algebraic property is assumed for $A$ and $B$ (other than the associative
property). Let us grade the terms of the commutator expansion of $\hat{f}$ by
the number of $[A,~]$ they carry. This is ambiguous. For instance
\begin{equation}
B^2 A = A B^2 -[A,B]B-B[A,B]
\label{eq:A.1}
.
\end{equation}
The expression as a whole is of zeroth order (this is the leading
order). However, the concrete zeroth, first and second order components are
different in the left and the right hand sides of the equation.

To remedy this situation, the ambiguity can be removed by choosing a canonical
form. A concrete choice comes from imposing the following prescriptions: i) In
the canonical form the expression is written as a linear combination of
products of blocks of the type, $A$ or $[A,~]^n B$ with $n=0,1,2,\ldots$ (that
is, $B$, $[A,B]$, $[A,[A,B]]$, \ldots). ii) The blocks $A$ are placed at the
left. Further, the $A$'s at the left count as order zero, and each block
$[A,~]^n B$ counts as order $n$. The right hand side of \eq{A.1} is written in
canonical form: the zeroth order is $A B^2$, the first order is
$-[A,B]B-B[A,B]$, and higher orders vanish.

Let us now show that the canonical form just defined, as well as the
corresponding grading of terms, can be derived from a bookkeeping parameter
using labeled operators. Namely, by counting powers of $\lambda$ in
\begin{equation}
\hat{f}=f(A,B) \to \hat{f}_\lambda=f(A_1+\lambda(A-A_1),B)
.
\label{eq:A.2}
\end{equation}
Here $A_1$ represents $A$ placed at the left (position 1 with respect to the
blocks $[A,~]^n B$). For instance
\begin{equation}
B^2 A \to B^2 (A_1+\lambda(A-A_1))
=
A B^2 + \lambda ( B^2 A - A B^2 )
=
A B^2 -\lambda ( [A,B]B +  B[A,B] )
.
\end{equation}

To proof \eq{A.2} in general, first note that
$A_i-A_{i+1}$ is just $[A,~]$ placed at position $i$. E.g., $(A_2-A_3)B^2
= A_2 B^2 - A_3 B^2 = BAB-B^2A = B[A,B]$. Then, if a block $A$ is located at
position $n$, one can write
\begin{equation}
A-A_1 = A_n-A_1 = (A_n -A_{n-1}) + (A_{n-1} - A_{n-2}) + \cdots + 
(A_2 - A_1)
.
\end{equation}
Therefore, $A-A_1$ is a sum of commutators and $\lambda$ in \eq{A.2} just
counts the number of commutators $[A,~]$.

This counting is unambiguous and extends trivially to the case of more
operators, $f(A,B,C,\ldots )$ if terms are still graded by the number of
$[A,~]$. It is worth noticing that things are more complicated for traced
expressions, due to the cyclic property of the trace. (For instance, position
``1'' becomes ambiguous.)

At the end of Section \ref{subsec:IIIA}, it was noted that choosing to put the
Polyakov loop at the left or at the right and then setting it to unity gives
different results and breaks hermiticity of the heat kernel at finite
temperature. This can be seen in the heat kernel coefficient $A_1$ in
\eq{3.26} since $\bar{\xi}_2$ is not zero for $\Omega=1$.  More generally,
consider \eq{2.38a} with $Q\to 0$. The covariant symbol $\bar{K}$ does not
break the symmetry but the momentum derivatives can only be taken to the right
due to the presence of $h_M(p_0)$ and this breaks the symmetry. Repeating the
calculation with $h_M(p_0)$ placed at the right gives a different result,
namely, the transposed of the previous one. The same conclusion can be
obtained from ordinary symbols. This can be illustrated with a simple
example. Consider the operator $(D_0+X)^{-1}$. We consider the two expansions,
with $D_0$ moved to the left or to the right:
\begin{equation}
\frac{1}{D_0+X}
=
  \frac{1}{D_{0,L}+X + D_0-D_{0,L}}
=
N_L-N_L(D_0-D_{0,L})N_L+\cdots
,
\end{equation}
here $N_L=(D_{0,L}+X)^{-1}$ and $D_{0,L}$ is $D_0$ at the left. If we set now
$D_{0,L}\to 0$ in $N_L$ the result is
\begin{equation}
N-N(D_0-D_{0,L})N+O(N^2)
=
N+[D_0,N]N+O(N^2)
,
\quad
N=X^{-1}
.
\end{equation}
A similar calculation with $D_{0,R}\to 0$ in $N_R$ gives
\begin{equation}
N-N(D_0-D_{0,R})N+O(N^2)
=
N-N[D_0,N]+O(N^2)
.
\end{equation}
So the two prescriptions differ by $[D_0,N^2]+O(N^2)$. On the other hand,
inside the trace, the two prescriptions do yield the same result (as stated in
\eq{4.38}).

\section{Momentum integrals at finite temperature}
\label{app:B}

Let
\begin{equation}
\langle p_{\mu_1} p_{\mu_2} \cdots p_{\mu_n} \rangle
=
(4\pi\tau)^{\dimst/2} \int\frac{d^{\dimst}p}{(2\pi)^{\dimst}}\,h_M(p_0-Q)
\,
e^{-\tau p^2} 
p_{\mu_1} p_{\mu_2} \cdots p_{\mu_n}
.
\end{equation}
These integrals are needed to obtain the heat kernel expansion coefficients at
finite temperature. They are not normalized to unity. In particular
\begin{equation}
\langle 1\rangle = \xi_0
.
\end{equation}

The basic result comes from distinguishing spatial from temporal degrees of
freedom:
\begin{eqnarray}
\langle p_{i_1} p_{i_2} \cdots p_{i_{2n}} p_0^m\rangle
&=&
(4\pi\tau)^{(\dims)/2}
\int\frac{d^{\dims}p}{(2\pi)^{\dims}}
e^{-\tau \vp^2} p_{i_1} p_{i_2} \cdots p_{i_{2n}}
(4\pi\tau)^{1/2}
\int\frac{dp_0}{2\pi}
\,h_M(p_0-Q) \, e^{-\tau p_0^2} p_0^m
\nonumber\\
&=&
\frac{1}{(2\tau)^n} 
\delta_{i_1 i_2\ldots i_{2n}}
\frac{1}{(i\sqrt{\tau})^m}
\varphi_m
.
\end{eqnarray}

Here the symbol $\delta_{i_1 i_2\ldots i_{2n}}$ represents the symmetric sum
of the $(2n-1)!!$ products of $n$ Kronecker deltas (each term with weight
one). E.g.
\begin{equation}
\delta_{ijkl} = 
\delta_{ij}\delta_{kl}+ \delta_{ik}\delta_{jl} + \delta_{il}\delta_{jk}
.
\end{equation}
Besides, we have introduced the auxiliary functions
\begin{equation}
  \varphi_m = (4\pi\tau)^{1/2} i^m \tau^{m/2}  \, \int\frac{dp_0}{2\pi}
\,h_M(p_0-Q) \,   e^{-\tau p_0^2}  p_0^m
,
\qquad
m=0,1,2,\ldots
\end{equation}
These are related to the functions $\xi_n$ of \eq{3.3} through the relations
\begin{equation}
  \varphi_m = \sum_{n=0}^m  i^{n+m} 2^{(n-m)/2} c^\prime_{nm}\,\xi_n
,
\qquad
\xi_n = \sum_{m=0}^n (-i)^{n+m} 2^{-(n-m)/2} c_{nm}  \, \varphi_m
,
\end{equation}
where
\begin{equation}
x^m = \sum_{n=0}^m c^\prime_{nm} H_n(x)
,
\qquad
 H_n(x) = \sum_{m=0}^n c_{nm} \, x^m
.
\end{equation}
As matrices $c^\prime = c^{-1T}$. 

In order to compute the heat kernel to four covariant derivatives, we need
$\langle p_{\mu_1} p_{\mu_2} \cdots p_{\mu_n} \rangle$ for $0 \le n\le 4$. Using
the previous formulas one obtains
\begin{eqnarray}
\langle 1\rangle &=& \xi_0
,
\nonumber\\
\langle p_{\mu}\rangle &=& \frac{i}{2 \tau^{1/2}}\delta_{\mu 0} \, \bar{\xi}_1
,
\nonumber\\
\langle p_\mu p_\nu\rangle &=& \frac{1}{2\tau}
(
\delta_{\mu\nu} \, \xi_0
-\frac{1}{2}\delta_{\mu 0} \delta_{\nu 0}\, \bar{\xi}_2 
)
,
\nonumber\\
\langle p_\mu p_\nu p_\alpha \rangle &=& \frac{i}{4\tau^{3/2}}
\left(
(
\delta_{\mu\nu}\delta_{\alpha 0}
+
\delta_{\mu\alpha}\delta_{\nu 0}
+
\delta_{\nu\alpha}\delta_{\mu 0}
) \bar{\xi}_1
-\frac{1}{2} \delta_{\mu 0} \delta_{\nu 0} \delta_{\alpha 0} \bar{\xi}_3
\right)
,
\nonumber\\
\langle p_\mu p_\nu p_\alpha p_\beta \rangle &=& \frac{1}{4\tau^2}
\left(
\delta_{\mu\nu\alpha\beta}\,\xi_0
-\frac{1}{2}(
\delta_{\mu\nu}\delta_{\alpha 0}\delta_{\beta 0}
+
\delta_{\mu\alpha}\delta_{\nu 0}\delta_{\beta 0}
+
\delta_{\mu\beta}\delta_{\nu 0}\delta_{\alpha 0}
+
\delta_{\nu\alpha}\delta_{\mu 0}\delta_{\beta 0}
\right.
\nonumber\\&&
\left.
+
\delta_{\nu\beta}\delta_{\mu 0}\delta_{\alpha 0}
+
\delta_{\alpha\beta}\delta_{\mu 0}\delta_{\nu 0}
)\, \bar{\xi}_2
+
\frac{1}{4} \delta_{\mu 0} \delta_{\nu 0} \delta_{\alpha 0}\delta_{\beta 0}
\bar{\xi}_4
\right)
,
\end{eqnarray}

The formulas in this appendix plus the first \eq{3.3} written as
\begin{equation}
\xi_n = (4\pi\tau)^{1/2} (-i)^n 2^{-n/2}  \, \int\frac{dp_0}{2\pi}
\,h_M(p_0-Q) \,   e^{-\tau p_0^2}  H_n(\sqrt{2\tau}p_0)
,
\end{equation}
hold if the function $h_M(p_0)$ is replaced everywhere by a more
general weight function, $h(p_0)$. No special property of $h_M(p_0)$
has been used.

\section{Formulas}
\label{app:C}

Covariant symbol of $K$ through fourth order in the derivative expansion:
\begin{eqnarray}
\bar{K}
&=&
X-p_{\mu}p_{\mu}
+iX_{\mu}\partial^p_{\mu}
+p_{\mu}F_{\mu\nu}\partial^p_{\nu
}-\frac{1}{2}X_{\mu\nu}\partial^p_{\mu}\partial^p_{\nu}
\nonumber\\&&
+\frac{2}{3}ip_{\mu}F_{\nu\mu\alpha}\partial^p_{\nu}\partial^p_{\alpha}
+\frac{1}{3}iF_{\mu\mu\nu}\partial^p_{\nu}
-\frac{1}{6}iX_{\mu\nu\alpha}\partial^p_{\mu}\partial^p_{\nu}\partial^p_{\alpha}
\nonumber\\&&
-\frac{1}{4}p_{\mu}F_{\nu\alpha\mu\beta}\partial^p_{\nu}\partial^p_{\alpha}\partial^p_{\beta}
-\frac{1}{4}F_{\mu\nu\mu\alpha}\partial^p_{\nu}\partial^p_{\alpha}
+\frac{1}{4}F_{\mu\nu}F_{\nu\alpha}\partial^p_{\mu}\partial^p_{\alpha}
+\frac{1}{24}X_{\mu\nu\alpha\beta}\partial^p_{\mu}\partial^p_{\nu}\partial^p_{\alpha}\partial^p_{\beta}
+O(D^5)
.
\end{eqnarray}

Diagonal matrix elements of the propagator through third order in the
derivative expansion, in $X$-form:
\begin{eqnarray}
\langle x|G(z)|x\rangle_h
&=&
\int\frac{d^{\dimst}p}{(2\pi)^{\dimst}}h(p+iD)\Big(
I_1 
-2 i I_{1,2} p_{\mu } X_{\mu}
-4 I_{1,3} p_{\mu } p_{\nu } X_{\mu \nu }
+I_{1,2} X_{\mu \mu }
\nonumber\\ &&
-\left(8 I_{1,1,3}+4I_{1,2,2}\right) p_{\mu } p_{\nu } X_{\mu } X_{\nu }
+2 I_{1,1,2} X_{\mu } X_{\mu }
-2i I_{1,1,2} p_{\mu } F_{\mu \nu } X_{\nu }
\nonumber\\ &&
-\frac{8}{3} i I_{1,3} p_{\mu } p_{\nu } p_{\alpha } F_{\mu \nu \alpha }
+\frac{8}{3} i I_{1,3} p_{\mu } p_{\nu } p_{\alpha } F_{\nu \mu \alpha }
-\frac{2}{3} i I_{1,2} p_{\mu } F_{\nu \mu \nu }
+8 i I_{1,4} p_{\mu } p_{\nu } p_{\alpha } X_{\mu \nu\alpha }
\nonumber\\ &&
-i \left(-24 I_{1,1,4}-8 I_{1,2,3}\right) p_{\mu } p_{\nu } p_{\alpha } X_{\mu} X_{\nu \alpha }
-i \left(-24 I_{1,1,4}-16 I_{1,2,3}
-8 I_{1,3,2}\right) p_{\mu } p_{\nu  } p_{\alpha } X_{\mu \nu } X_{\alpha }
\nonumber \\ &&
-i \left(-48 I_{1,1,1,4}-32 I_{1,1,2,3}-16 I_{1,1,3,2}-16 I_{1,2,1,3}-8 I_{1,2,2,2}\right) p_{\mu } p_{\nu } p_{\alpha } X_{\mu} X_{\nu } X_{\alpha }
-\frac{4}{3} i I_{1,3} p_{\mu } X_{\mu \nu \nu }
\nonumber \\ &&
-\frac{4}{3} i I_{1,3} p_{\mu } X_{\nu \mu \nu }
-\frac{4}{3} i I_{1,3} p_{\mu } X_{\nu \nu \mu }
+i \left(-4 I_{1,1,3}-2 I_{1,2,2}\right) p_{\mu } X_{\mu } X_{\nu \nu }
-4 i I_{1,1,3} p_{\mu } X_{\nu } X_{\mu \nu }
\nonumber\\ &&
-4 i I_{1,1,3} p_{\mu } X_{\nu } X_{\nu \mu }
+i \left(-4I_{1,1,3}-2 I_{1,2,2}\right) p_{\mu } X_{\mu \nu } X_{\nu }
+i \left(-4 I_{1,1,3}-2 I_{1,2,2}\right) p_{\mu } X_{\nu \mu } X_{\nu }
\nonumber\\ &&
+i \left(-4 I_{1,1,3}-2 I_{1,2,2}\right) p_{\mu } X_{\nu \nu } X_{\mu }
+i \left(-8 I_{1,1,1,3}-4 I_{1,1,2,2}-4 I_{1,2,1,2}\right) p_{\mu } X_{\mu } X_{\nu } X_{\nu }
\nonumber \\ &&
+i \left(-8 I_{1,1,1,3}-4 I_{1,1,2,2}\right) p_{\mu } X_{\nu } X_{\mu } X_{\nu }
+i \left(-8 I_{1,1,1,3}-4 I_{1,1,2,2}\right) p_{\mu } X_{\nu } X_{\nu } X_{\mu }
+O(D^4)
\Big)
.
\label{eq:4.12a}
\end{eqnarray}

Covariant symbol of $G(z)$ in $N$-form, through third order:
\begin{eqnarray}
\bar{G}(z)
&=&
N
+iN_{\mu}\partial^p_{\mu}
-2ip_{\mu}N_{\mu}N
\nonumber\\&&
+N_{\mu\mu}N
-\frac{1}{2}N_{\mu\nu}\partial^p_{\mu}\partial^p_{\nu}
+2p_{\mu}N_{\mu}N_{\nu}\partial^p_{\nu}
+p_{\mu}N_{\mu\nu}N\partial^p_{\nu}
+p_{\mu}N_{\nu\mu}N\partial^p_{\nu}
\nonumber\\&&
+p_{\mu}NF_{\mu\nu}N\partial^p_{\nu}
-4p_{\mu}p_{\nu}N_{\mu}N_{\nu}N
-4p_{\mu}p_{\nu}N_{\mu\nu}N^2
\nonumber\\&&
+iN_{\mu\mu}N_{\nu}\partial^p_{\nu}
+\frac{1}{3}iN_{\mu\mu\nu}N\partial^p_{\nu}
+\frac{1}{3}iN_{\mu\nu\mu}N\partial^p_{\nu}
+\frac{1}{3}iN_{\mu\nu\nu}N\partial^p_{\mu}
+\frac{1}{3}iNF_{\mu\mu\nu}N\partial^p_{\nu}
\nonumber\\&&
+iN_{\mu}F_{\mu\nu}N\partial^p_{\nu}
-\frac{1}{6}iN_{\mu\nu\alpha}\partial^p_{\mu}\partial^p_{\nu}\partial^p_{\alpha}
-2ip_{\mu}N_{\mu}N_{\nu\nu}N
-2ip_{\mu}N_{\mu\nu}N_{\nu}N
\nonumber\\&&
-2ip_{\mu}N_{\nu\mu}N_{\nu}N
-2ip_{\mu}N_{\nu\nu}N_{\mu}N
-\frac{4}{3}ip_{\mu}N_{\mu\nu\nu}N^2
-\frac{4}{3}ip_{\mu}N_{\nu\mu\nu}N^2
-\frac{4}{3}ip_{\mu}N_{\nu\nu\mu}N^2
\nonumber\\&&
-2ip_{\mu}NF_{\mu\nu}N_{\nu}N
-\frac{2}{3}ip_{\mu}NF_{\nu\mu\nu}N^2
+ip_{\mu}N_{\mu}N_{\nu\alpha}\partial^p_{\nu}\partial^p_{\alpha}
+ip_{\mu}N_{\mu\nu}N_{\alpha}\partial^p_{\nu}\partial^p_{\alpha}
\nonumber\\&&
+ip_{\mu}N_{\nu\mu}N_{\alpha}\partial^p_{\nu}\partial^p_{\alpha}
+\frac{1}{3}ip_{\mu}N_{\mu\nu\alpha}N\partial^p_{\nu}\partial^p_{\alpha}
+\frac{1}{3}ip_{\mu}N_{\nu\mu\alpha}N\partial^p_{\nu}\partial^p_{\alpha}
+\frac{1}{3}ip_{\mu}N_{\nu\alpha\mu}N\partial^p_{\nu}\partial^p_{\alpha}
\nonumber\\&&
+ip_{\mu}NF_{\mu\nu}N_{\alpha}\partial^p_{\nu}\partial^p_{\alpha}
+\frac{2}{3}ip_{\mu}NF_{\nu\mu\alpha}N\partial^p_{\nu}\partial^p_{\alpha}
+ip_{\mu}N_{\nu}F_{\mu\alpha}N\partial^p_{\nu}\partial^p_{\alpha}
-4ip_{\mu}p_{\nu}N_{\mu}N_{\nu}N_{\alpha}\partial^p_{\alpha}
\nonumber\\&&
-2ip_{\mu}p_{\nu}N_{\mu}N_{\nu\alpha}N\partial^p_{\alpha}
-2ip_{\mu}p_{\nu}N_{\mu}N_{\alpha\nu}N\partial^p_{\alpha}
-4ip_{\mu}p_{\nu}N_{\mu\nu}NN_{\alpha}\partial^p_{\alpha}
-4ip_{\mu}p_{\nu}N_{\mu\nu}N_{\alpha}N\partial^p_{\alpha}
\nonumber\\&&
-2ip_{\mu}p_{\nu}N_{\mu\alpha}N_{\nu}N\partial^p_{\alpha}
-2ip_{\mu}p_{\nu}N_{\alpha\mu}N_{\nu}N\partial^p_{\alpha}
-\frac{4}{3}ip_{\mu}p_{\nu}N_{\mu\nu\alpha}N^2\partial^p_{\alpha}
-\frac{4}{3}ip_{\mu}p_{\nu}N_{\mu\alpha\nu}N^2\partial^p_{\alpha}
\nonumber\\&&
-\frac{4}{3}ip_{\mu}p_{\nu}N_{\alpha\mu\nu}N^2\partial^p_{\alpha}
-2ip_{\mu}p_{\nu}NF_{\mu\alpha}N_{\nu}N\partial^p_{\alpha}
-\frac{4}{3}ip_{\mu}p_{\nu}NF_{\mu\nu\alpha}N^2\partial^p_{\alpha}
-2ip_{\mu}p_{\nu}N_{\mu}F_{\nu\alpha}N^2\partial^p_{\alpha}
\nonumber\\&&
-2ip_{\mu}p_{\nu}N_{\mu}NF_{\nu\alpha}N\partial^p_{\alpha}
+8ip_{\mu}p_{\nu}p_{\alpha}N_{\mu}N_{\nu}N_{\alpha}N
+8ip_{\mu}p_{\nu}p_{\alpha}N_{\mu}N_{\nu\alpha}N^2
\nonumber\\&&
+8ip_{\mu}p_{\nu}p_{\alpha}N_{\mu\nu}NN_{\alpha}N
+16ip_{\mu}p_{\nu}p_{\alpha}N_{\mu\nu}N_{\alpha}N^2
+8ip_{\mu}p_{\nu}p_{\alpha}N_{\mu\nu\alpha}N^3
+O(D^4)
.
\end{eqnarray}

Fourth order of the diagonal matrix element of $G(z)$ in $N$-form:
\begin{eqnarray}
\langle x|G(z)|x\rangle_{h,4}
&=&
\int\frac{d^{\dimst}p}{(2\pi)^{\dimst}}h(p+iD)\Big(
\nonumber\\&&
N_{\mu\mu}N_{\nu\nu}N
+\frac{2}{3}N_{\mu\mu\nu}N_{\nu}N
+\frac{2}{3}N_{\mu\nu\mu}N_{\nu}N
+\frac{2}{3}N_{\mu\nu\nu}N_{\mu}N
+\frac{1}{3}N_{\mu\mu\nu\nu}N^2
+\frac{1}{3}N_{\mu\nu\mu\nu}N^2
\nonumber\\&&
+\frac{1}{3}N_{\mu\nu\nu\mu}N^2
+\frac{2}{3}NF_{\mu\mu\nu}N_{\nu}N
+2N_{\mu}F_{\mu\nu}N_{\nu}N
+\frac{2}{3}N_{\mu}F_{\nu\mu\nu}N^2
+\frac{1}{2}NF_{\mu\nu}F_{\mu\nu}N^2
\nonumber\\&&
-4p_{\mu}p_{\nu}N_{\mu}N_{\nu}N_{\alpha\alpha}N
-4p_{\mu}p_{\nu}N_{\mu}N_{\nu\alpha}N_{\alpha}N
-4p_{\mu}p_{\nu}N_{\mu}N_{\alpha\nu}N_{\alpha}N
-4p_{\mu}p_{\nu}N_{\mu}N_{\alpha\alpha}N_{\nu}N
\nonumber\\&&
-\frac{8}{3}p_{\mu}p_{\nu}N_{\mu}N_{\nu\alpha\alpha}N^2
-\frac{8}{3}p_{\mu}p_{\nu}N_{\mu}N_{\alpha\nu\alpha}N^2
-\frac{8}{3}p_{\mu}p_{\nu}N_{\mu}N_{\alpha\alpha\nu}N^2
-4p_{\mu}p_{\nu}N_{\mu\nu}NN_{\alpha\alpha}N
\nonumber\\&&
-8p_{\mu}p_{\nu}N_{\mu\nu}N_{\alpha}N_{\alpha}N
-8p_{\mu}p_{\nu}N_{\mu\nu}N_{\alpha\alpha}N^2
-4p_{\mu}p_{\nu}N_{\mu\alpha}N_{\nu}N_{\alpha}N
-4p_{\mu}p_{\nu}N_{\mu\alpha}N_{\alpha}N_{\nu}N
\nonumber\\&&
-4p_{\mu}p_{\nu}N_{\mu\alpha}N_{\nu\alpha}N^2
-4p_{\mu}p_{\nu}N_{\mu\alpha}N_{\alpha\nu}N^2
-4p_{\mu}p_{\nu}N_{\alpha\mu}N_{\nu}N_{\alpha}N
-4p_{\mu}p_{\nu}N_{\alpha\mu}N_{\alpha}N_{\nu}N
\nonumber\\&&
-4p_{\mu}p_{\nu}N_{\alpha\mu}N_{\nu\alpha}N^2
-4p_{\mu}p_{\nu}N_{\alpha\mu}N_{\alpha\nu}N^2
-4p_{\mu}p_{\nu}N_{\alpha\alpha}N_{\mu}N_{\nu}N
-4p_{\mu}p_{\nu}N_{\alpha\alpha}N_{\mu\nu}N^2
\nonumber\\&&
-\frac{8}{3}p_{\mu}p_{\nu}N_{\mu\nu\alpha}NN_{\alpha}N
-\frac{16}{3}p_{\mu}p_{\nu}N_{\mu\nu\alpha}N_{\alpha}N^2
-\frac{8}{3}p_{\mu}p_{\nu}N_{\mu\alpha\nu}NN_{\alpha}N
-\frac{16}{3}p_{\mu}p_{\nu}N_{\mu\alpha\nu}N_{\alpha}N^2
\nonumber\\&&
-\frac{8}{3}p_{\mu}p_{\nu}N_{\mu\alpha\alpha}NN_{\nu}N
-\frac{16}{3}p_{\mu}p_{\nu}N_{\mu\alpha\alpha}N_{\nu}N^2
-\frac{8}{3}p_{\mu}p_{\nu}N_{\alpha\mu\nu}NN_{\alpha}N
-\frac{16}{3}p_{\mu}p_{\nu}N_{\alpha\mu\nu}N_{\alpha}N^2
\nonumber\\&&
-\frac{8}{3}p_{\mu}p_{\nu}N_{\alpha\mu\alpha}NN_{\nu}N
-\frac{16}{3}p_{\mu}p_{\nu}N_{\alpha\mu\alpha}N_{\nu}N^2
-\frac{8}{3}p_{\mu}p_{\nu}N_{\alpha\alpha\mu}NN_{\nu}N
-\frac{16}{3}p_{\mu}p_{\nu}N_{\alpha\alpha\mu}N_{\nu}N^2
\nonumber\\&&
-2p_{\mu}p_{\nu}N_{\mu\nu\alpha\alpha}N^3
-2p_{\mu}p_{\nu}N_{\mu\alpha\nu\alpha}N^3
-2p_{\mu}p_{\nu}N_{\mu\alpha\alpha\nu}N^3
-2p_{\mu}p_{\nu}N_{\alpha\mu\nu\alpha}N^3
\nonumber\\&&
-2p_{\mu}p_{\nu}N_{\alpha\mu\alpha\nu}N^3
-2p_{\mu}p_{\nu}N_{\alpha\alpha\mu\nu}N^3
-4p_{\mu}p_{\nu}NF_{\mu\alpha}N_{\nu}N_{\alpha}N
-4p_{\mu}p_{\nu}NF_{\mu\alpha}N_{\alpha}N_{\nu}N
\nonumber\\&&
-4p_{\mu}p_{\nu}NF_{\mu\alpha}N_{\nu\alpha}N^2
-4p_{\mu}p_{\nu}NF_{\mu\alpha}N_{\alpha\nu}N^2
-\frac{8}{3}p_{\mu}p_{\nu}NF_{\mu\nu\alpha}NN_{\alpha}N
-\frac{16}{3}p_{\mu}p_{\nu}NF_{\mu\nu\alpha}N_{\alpha}N^2
\nonumber\\&&
-\frac{4}{3}p_{\mu}p_{\nu}NF_{\alpha\mu\alpha}NN_{\nu}N
-\frac{8}{3}p_{\mu}p_{\nu}NF_{\alpha\mu\alpha}N_{\nu}N^2
-p_{\mu}p_{\nu}NF_{\mu\alpha\nu\alpha}N^3
-p_{\mu}p_{\nu}NF_{\alpha\mu\nu\alpha}N^3
\nonumber\\&&
-4p_{\mu}p_{\nu}N_{\mu}F_{\nu\alpha}NN_{\alpha}N
-8p_{\mu}p_{\nu}N_{\mu}F_{\nu\alpha}N_{\alpha}N^2
-\frac{8}{3}p_{\mu}p_{\nu}N_{\mu}F_{\alpha\nu\alpha}N^3
-4p_{\mu}p_{\nu}N_{\mu}NF_{\nu\alpha}N_{\alpha}N
\nonumber\\&&
-\frac{4}{3}p_{\mu}p_{\nu}N_{\mu}NF_{\alpha\nu\alpha}N^2
-2p_{\mu}p_{\nu}NF_{\mu\alpha}F_{\nu\alpha}N^3
+16p_{\mu}p_{\nu}p_{\alpha}p_{\beta}N_{\mu}N_{\nu}N_{\alpha}N_{\beta}N
\nonumber\\&&
+16p_{\mu}p_{\nu}p_{\alpha}p_{\beta}N_{\mu}N_{\nu}N_{\alpha\beta}N^2
+16p_{\mu}p_{\nu}p_{\alpha}p_{\beta}N_{\mu}N_{\nu\alpha}NN_{\beta}N
+32p_{\mu}p_{\nu}p_{\alpha}p_{\beta}N_{\mu}N_{\nu\alpha}N_{\beta}N^2
\nonumber\\&&
+16p_{\mu}p_{\nu}p_{\alpha}p_{\beta}N_{\mu}N_{\nu\alpha\beta}N^3
+16p_{\mu}p_{\nu}p_{\alpha}p_{\beta}N_{\mu\nu}NN_{\alpha}N_{\beta}N
+16p_{\mu}p_{\nu}p_{\alpha}p_{\beta}N_{\mu\nu}NN_{\alpha\beta}N^2
\nonumber\\&&
+32p_{\mu}p_{\nu}p_{\alpha}p_{\beta}N_{\mu\nu}N_{\alpha}NN_{\beta}N
+64p_{\mu}p_{\nu}p_{\alpha}p_{\beta}N_{\mu\nu}N_{\alpha}N_{\beta}N^2
+48p_{\mu}p_{\nu}p_{\alpha}p_{\beta}N_{\mu\nu}N_{\alpha\beta}N^3
\nonumber\\&&
+16p_{\mu}p_{\nu}p_{\alpha}p_{\beta}N_{\mu\nu\alpha}N^2N_{\beta}N
+32p_{\mu}p_{\nu}p_{\alpha}p_{\beta}N_{\mu\nu\alpha}NN_{\beta}N^2
+48p_{\mu}p_{\nu}p_{\alpha}p_{\beta}N_{\mu\nu\alpha}N_{\beta}N^3
\nonumber\\&&
+16p_{\mu}p_{\nu}p_{\alpha}p_{\beta}N_{\mu\nu\alpha\beta}N^4
\Big)
.
\end{eqnarray}

\section{The cyclic property in $h$-spaces}
\label{app:D}

In order to prove \eq{4.16}, we can assume, without loss of generality, that
$A(p)=\hat{A} \, a(p)$ and $B(p)=\hat{B}\, b(p)$, where the operators
$\hat{A}$ and $\hat{B}$ do not depend on $p_\mu$ and $a(p)$ and $b(p)$ are
c-numbers (i.e., they commute with everything, except $\partial^p_\mu$).

\begin{eqnarray}
\int\frac{d^{\dimst}p}{(2\pi)^{\dimst}}
A(p)h(p+iD) B(p)
&=&
\int\frac{d^{\dimst}p}{(2\pi)^{\dimst}}
\hat{A}\,h(p+iD) \hat{B}\,a(p)\,b(p)
\nonumber \\ 
&=&
\int\frac{d^{\dimst}p}{(2\pi)^{\dimst}}
\hat{A}\,h(p) e^{-i\partial^p D}\hat{B}\,a(p)\,b(p)
\nonumber \\ 
&=&
\int\frac{d^{\dimst}p}{(2\pi)^{\dimst}}
\,h(p)\, \hat{A}\, e^{-i\partial^p D}\hat{B}\,a(p)\,b(p)
\nonumber \\ 
&=&
\int\frac{d^{\dimst}p}{(2\pi)^{\dimst}}
\,h(p+iD)\, e^{i\partial^p D}\, \hat{A}\, e^{-i\partial^p D}\hat{B}\,a(p)\,b(p)
\nonumber \\ 
&=&
\int\frac{d^{\dimst}p}{(2\pi)^{\dimst}}
h(p+iD) \, e^{i\partial^p\hat{D}_A} \hat{A}\,\hat{B}\,a(p)\,b(p)
\nonumber \\ 
&=&
\int\frac{d^{\dimst}p}{(2\pi)^{\dimst}}
h(p+iD) \,e^{i\partial^p\hat{D}_A} A(p) B(p)
.
\end{eqnarray}


\end{document}